\pdfoutput=1

\documentclass[12pt,a4paper]{article}
\usepackage{ifthen} 
\newboolean{pdflatex}
\setboolean{pdflatex}{true} 

\newboolean{articletitles}
\setboolean{articletitles}{true} 

\newboolean{uprightparticles}
\setboolean{uprightparticles}{false} 

\newboolean{inbibliography}
\setboolean{inbibliography}{false} 

\usepackage[normalem]{ulem}

\usepackage[top=1in, bottom=1.25in, left=1in, right=1in]{geometry}

\usepackage{microtype}
\usepackage{lineno}  
\usepackage{xspace} 
\usepackage{caption} 

\usepackage{graphicx}  
\usepackage{color}
\usepackage{colortbl}
\graphicspath{{./}} 

\usepackage{amsmath} 
\usepackage{amssymb}
\usepackage{amsfonts}
\usepackage{upgreek} 

\newcommand*\patchAmsMathEnvironmentForLineno[1]{%
\expandafter\let\csname old#1\expandafter\endcsname\csname #1\endcsname
\expandafter\let\csname oldend#1\expandafter\endcsname\csname
end#1\endcsname
 \renewenvironment{#1}%
   {\linenomath\csname old#1\endcsname}%
   {\csname oldend#1\endcsname\endlinenomath}%
}
\newcommand*\patchBothAmsMathEnvironmentsForLineno[1]{%
  \patchAmsMathEnvironmentForLineno{#1}%
  \patchAmsMathEnvironmentForLineno{#1*}%
}
\AtBeginDocument{%
\patchBothAmsMathEnvironmentsForLineno{equation}%
\patchBothAmsMathEnvironmentsForLineno{align}%
\patchBothAmsMathEnvironmentsForLineno{flalign}%
\patchBothAmsMathEnvironmentsForLineno{alignat}%
\patchBothAmsMathEnvironmentsForLineno{gather}%
\patchBothAmsMathEnvironmentsForLineno{multline}%
\patchBothAmsMathEnvironmentsForLineno{eqnarray}%
}

\usepackage{hyperref}    
\usepackage[all]{hypcap} 

\usepackage{xspace} 
\usepackage{upgreek}

\def\lhcb {\mbox{LHCb}\xspace}

\def\MagUp {\mbox{\em Mag\kern -0.05em Up}\xspace}

\ifthenelse{\boolean{uprightparticles}}%
{

 \def\Ppi         {\ensuremath{\uppi}\xspace}                 
                  
 \def\Prho        {\ensuremath{\uprho}\xspace}

 \def\PDelta      {\ensuremath{\Delta}\xspace}                 
 \def\PXi      {\ensuremath{\Xi}\xspace}                 
 \def\PLambda      {\ensuremath{\Lambda}\xspace}                 
 \def\PSigma      {\ensuremath{\Sigma}\xspace}                 
 \def\POmega      {\ensuremath{\Omega}\xspace}                 
 \def\PUpsilon      {\ensuremath{\Upsilon}\xspace}                 
 
 \def\PB      {\ensuremath{\mathrm{B}}\xspace}                 
                  
 \def\PD      {\ensuremath{\mathrm{D}}\xspace}

 \def\PK      {\ensuremath{\mathrm{K}}\xspace}

 \def\Pb      {\ensuremath{\mathrm{b}}\xspace}                 
 \def\Pc      {\ensuremath{\mathrm{c}}\xspace}                 
 \def\Pd      {\ensuremath{\mathrm{d}}\xspace}

 \def\Pi      {\ensuremath{\mathrm{i}}\xspace}

 \def\Pu      {\ensuremath{\mathrm{u}}\xspace}

}
{

 \def\Ppi         {\ensuremath{\pi}\xspace}                 
                  
 \def\Prho        {\ensuremath{\rho}\xspace}

 \mathchardef\PDelta="7101
 \mathchardef\PXi="7104
 \mathchardef\PLambda="7103
 \mathchardef\PSigma="7106
 \mathchardef\POmega="710A
 \mathchardef\PUpsilon="7107
                  
 \def\PB      {\ensuremath{B}\xspace}                 
                  
 \def\PD      {\ensuremath{D}\xspace}

 \def\PK      {\ensuremath{K}\xspace}

 \def\Pb      {\ensuremath{b}\xspace}                 
 \def\Pc      {\ensuremath{c}\xspace}                 
 \def\Pd      {\ensuremath{d}\xspace}

 \def\Pi      {\ensuremath{i}\xspace}

 \def\Pu      {\ensuremath{u}\xspace}

}

\makeatletter
\ifcase \@ptsize \relax
  \newcommand{\miniscule}{\@setfontsize\miniscule{4}{5}}
\or
  \newcommand{\miniscule}{\@setfontsize\miniscule{5}{6}}
\or
  \newcommand{\miniscule}{\@setfontsize\miniscule{5}{6}}
\fi
\makeatother

\DeclareRobustCommand{\optbar}[1]{\shortstack{{\miniscule (\rule[.5ex]{1.25em}{.18mm})}
  \\ [-.7ex] $#1$}}

\def\uquark    {{\ensuremath{\Pu}}\xspace}

\def\dquark    {{\ensuremath{\Pd}}\xspace}
\def\dquarkbar {{\ensuremath{\overline \dquark}}\xspace}

\def\cquark    {{\ensuremath{\Pc}}\xspace}

\def\bquark    {{\ensuremath{\Pb}}\xspace}

\def\pion   {{\ensuremath{\Ppi}}\xspace}
\def\piz    {{\ensuremath{\pion^0}}\xspace}

\def\pip    {{\ensuremath{\pion^+}}\xspace}
\def\pim    {{\ensuremath{\pion^-}}\xspace}
\def\pipm   {{\ensuremath{\pion^\pm}}\xspace}

\def\rhomeson {{\ensuremath{\Prho}}\xspace}
\def\rhoz     {{\ensuremath{\rhomeson^0}}\xspace}

\def\kaon    {{\ensuremath{\PK}}\xspace}
  \def\Kbar    {{\kern 0.2em\overline{\kern -0.2em \PK}{}}\xspace}

\def\KorKbar    {\kern 0.18em\optbar{\kern -0.18em K}{}\xspace}

\def\Kp      {{\ensuremath{\kaon^+}}\xspace}
\def\Km      {{\ensuremath{\kaon^-}}\xspace}
\def\Kpm     {{\ensuremath{\kaon^\pm}}\xspace}

\def\KS      {{\ensuremath{\kaon^0_{\mathrm{ \scriptscriptstyle S}}}}\xspace}

  \def\Dbar    {{\kern 0.2em\overline{\kern -0.2em \PD}{}}\xspace}
\def\D       {{\ensuremath{\PD}}\xspace}

\def\DorDbar    {\kern 0.18em\optbar{\kern -0.18em D}{}\xspace}
\def\Dz      {{\ensuremath{\D^0}}\xspace}
\def\Dzb     {{\ensuremath{\Dbar{}^0}}\xspace}

\def\Dstar   {{\ensuremath{\D^*}}\xspace}

\def\Dstarp  {{\ensuremath{\D^{*+}}}\xspace}
\def\Dstarm  {{\ensuremath{\D^{*-}}}\xspace}

\def\Bbar    {{\ensuremath{\kern 0.18em\overline{\kern -0.18em \PB}{}}}\xspace}

\def\BorBbar    {\kern 0.18em\optbar{\kern -0.18em B}{}\xspace}

  \def\Y#1S{\ensuremath{\PUpsilon{(#1S)}}\xspace}

\def\Lbar        {{\ensuremath{\kern 0.1em\overline{\kern -0.1em\PLambda}}}\xspace}
\def\LorLbar    {\kern 0.18em\optbar{\kern -0.18em \PLambda}{}\xspace}

\newcommand{\decay}[2]{\ensuremath{#1\!\to #2}\xspace}         

\def\to                 {\ensuremath{\rightarrow}\xspace}

\def\CP                {{\ensuremath{C\!P}}\xspace}

\newcommand{\dm}{{\ensuremath{\Delta m}}\xspace}

\def\AT#1     {\ensuremath{A_{\mathrm{T}}^{#1}}\xspace}           

\def\C#1      {\ensuremath{\mathcal{C}_{#1}}\xspace}                       
\def\Cp#1     {\ensuremath{\mathcal{C}_{#1}^{'}}\xspace}                    
\def\Ceff#1   {\ensuremath{\mathcal{C}_{#1}^{\mathrm{(eff)}}}\xspace}        
\def\Cpeff#1  {\ensuremath{\mathcal{C}_{#1}^{'\mathrm{(eff)}}}\xspace}       
\def\Ope#1    {\ensuremath{\mathcal{O}_{#1}}\xspace}                       
\def\Opep#1   {\ensuremath{\mathcal{O}_{#1}^{'}}\xspace}                    

\newcommand{\tev}{\ifthenelse{\boolean{inbibliography}}{\ensuremath{~T\kern -0.05em eV}\xspace}{\ensuremath{\mathrm{\,Te\kern -0.1em V}}}\xspace}
\newcommand{\gev}{\ensuremath{\mathrm{\,Ge\kern -0.1em V}}\xspace}
\newcommand{\mev}{\ensuremath{\mathrm{\,Me\kern -0.1em V}}\xspace}
\newcommand{\kev}{\ensuremath{\mathrm{\,ke\kern -0.1em V}}\xspace}
\newcommand{\ev}{\ensuremath{\mathrm{\,e\kern -0.1em V}}\xspace}
\newcommand{\gevc}{\ensuremath{{\mathrm{\,Ge\kern -0.1em V\!/}c}}\xspace}
\newcommand{\mevc}{\ensuremath{{\mathrm{\,Me\kern -0.1em V\!/}c}}\xspace}
\newcommand{\gevcc}{\ensuremath{{\mathrm{\,Ge\kern -0.1em V\!/}c^2}}\xspace}
\newcommand{\gevgevcccc}{\ensuremath{{\mathrm{\,Ge\kern -0.1em V^2\!/}c^4}}\xspace}
\newcommand{\mevcc}{\ensuremath{{\mathrm{\,Me\kern -0.1em V\!/}c^2}}\xspace}

\def\mum  {\ensuremath{{\,\upmu\mathrm{m}}}\xspace}

\def\invfb   {\ensuremath{\mbox{\,fb}^{-1}}\xspace}

\newcommand{\chisq}{\ensuremath{\chi^2}\xspace}

\def\gsim{{~\raise.15em\hbox{$>$}\kern-.85em
          \lower.35em\hbox{$\sim$}~}\xspace}
\def\lsim{{~\raise.15em\hbox{$<$}\kern-.85em
          \lower.35em\hbox{$\sim$}~}\xspace}

\def\ptot       {\mbox{$p$}\xspace}
\def\pt         {\mbox{$p_{\mathrm{ T}}$}\xspace}

\def\evtgen     {\mbox{\textsc{EvtGen}}\xspace}

\def\geant      {\mbox{\textsc{Geant4}}\xspace}

\def\pythia     {\mbox{\textsc{Pythia}}\xspace}

\def\tell1  {TELL1\xspace}
\def\ukl1   {UKL1\xspace}

\newcommand{\eg}{\mbox{\itshape e.g.}\xspace}
\newcommand{\ie}{\mbox{\itshape i.e.}\xspace}

\usepackage{cite} 
\usepackage{mciteplus}

\def\pvalue     {$p$-value\xspace}
\def\pvalues    {$p$-values\xspace}

\begin{document}

\renewcommand{\thefootnote}{\fnsymbol{footnote}}
\setcounter{footnote}{1}


\begin{titlepage}
\pagenumbering{roman}

\vspace*{-1.5cm}
\centerline{\large EUROPEAN ORGANIZATION FOR NUCLEAR RESEARCH (CERN)}
\vspace*{1.5cm}
\noindent
\begin{tabular*}{\linewidth}{lc@{\extracolsep{\fill}}r@{\extracolsep{0pt}}}
\ifthenelse{\boolean{pdflatex}}
{\vspace*{-2.7cm}\mbox{\!\!\!\includegraphics[width=.14\textwidth]{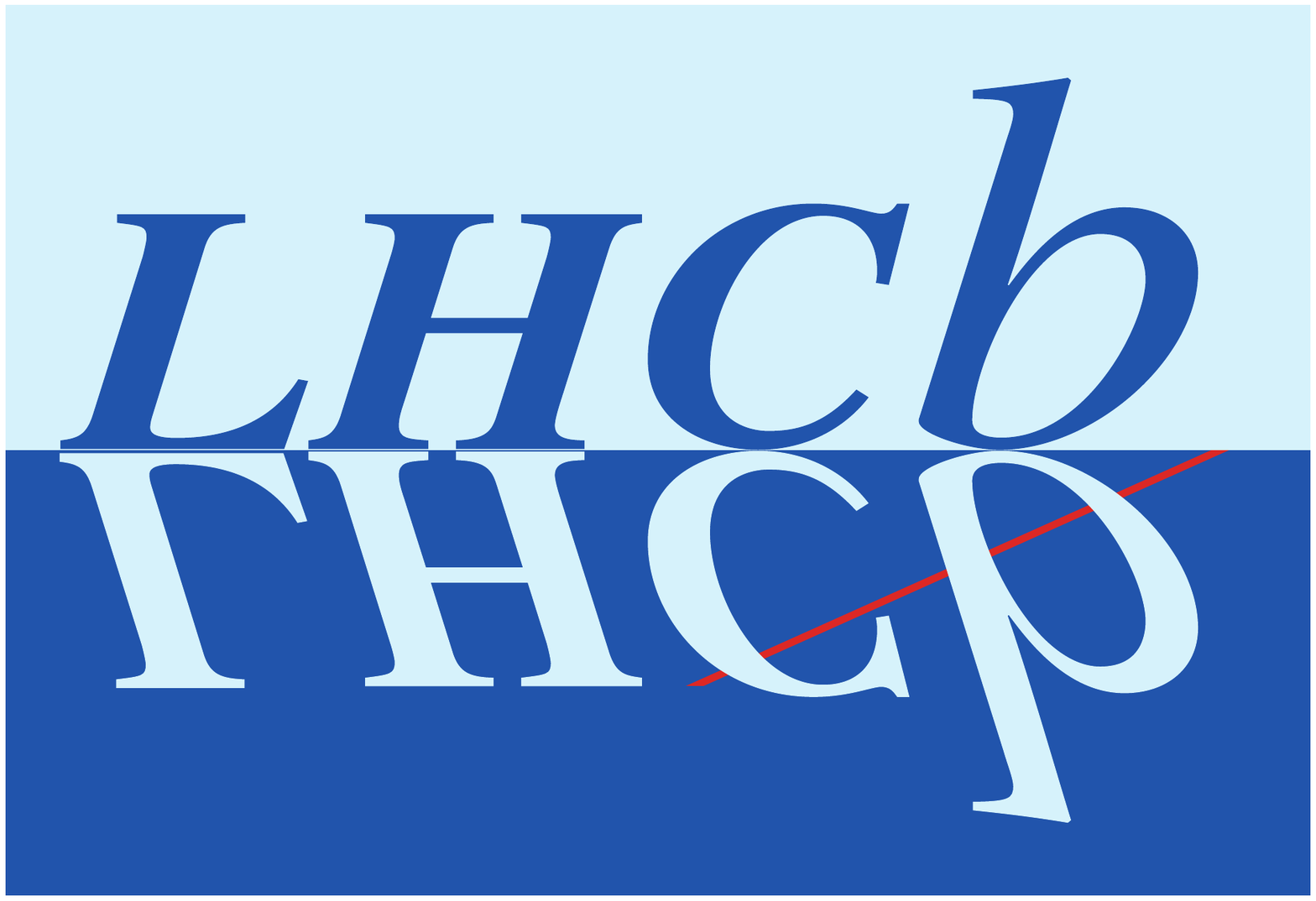}} & &}%
{\vspace*{-1.2cm}\mbox{\!\!\!\includegraphics[width=.12\textwidth]{lhcb-logo.eps}} & &}%
\\
 & & CERN-EP-2016-287 \\  
 & & LHCb-PAPER-2016-044  \\  
 & & 4 April 2017 \\ 
 
\end{tabular*}

\vspace*{4.0cm}

{\normalfont\bfseries\boldmath\huge
\begin{center}
  Search for $CP$ violation in the phase space of $D^0\rightarrow\pi^+\pi^-\pi^+\pi^-$ decays
\end{center}
}

\vspace*{2.0cm}

\begin{center}
The LHCb collaboration\footnote{Authors are listed at the end of this Letter.}
\end{center}

\vspace{\fill}

\begin{abstract}
  \noindent
A search for time-integrated $CP$ violation in the Cabibbo-suppressed
decay  \mbox{$D^0\rightarrow\pi^+\pi^-\pi^+\pi^-$} is performed using an unbinned, 
model-independent technique known as the energy test. This is the
first application of the energy test in four-body decays. The search is
performed for $P$-even  $CP$ asymmetries and, for the first time, 
is extended to probe the $P$-odd case. 
Using proton-proton collision data corresponding to an integrated
luminosity of 3.0~fb$^{-1}$ collected by the LHCb detector at 
centre-of-mass energies of $\sqrt{s}=$7~TeV and 8~TeV, the world's best sensitivity 
to $CP$ violation in this decay is obtained.  The data are found to be
consistent with the hypothesis of $CP$ symmetry with a $p$-value of
$(4.6\pm0.5)\%$ in the $P$-even case, and marginally consistent with
a $p$-value of $(0.6\pm0.2)\%$ in the $P$-odd 
case,  corresponding to a significance for $CP$ non-conservation
of 2.7 standard deviations. 

\end{abstract}

\vspace*{2.0cm}

\begin{center}
  Published in Phys.~Lett.~B~769~(2017)~345-356
\end{center}

\vspace{\fill}

{\footnotesize 
\centerline{\copyright~CERN on behalf of the LHCb collaboration, licence \href{http://creativecommons.org/licenses/by/4.0/}{CC-BY-4.0}.}}
\vspace*{2mm}

\end{titlepage}


\newpage
\setcounter{page}{2}
\mbox{~}

\cleardoublepage

\renewcommand{\thefootnote}{\arabic{footnote}}
\setcounter{footnote}{0}


\pagestyle{plain} 
\setcounter{page}{1}
\pagenumbering{arabic}

\section{Introduction}
\label{sec:Introduction}

The decay \decay{\Dz}{\pip\pim\pip\pim} (charge-conjugate decays are implied unless stated otherwise) 
proceeds via a singly Cabibbo-suppressed \decay{\cquark}{\dquark\uquark\dquarkbar} 
transition with an admixture from a \decay{\cquark}{\uquark g} gluonic-penguin transition.
Within the Standard Model (SM), these amplitudes have different weak phases,  and the interference between 
them may give rise to a violation of the charge-parity (\CP) symmetry in the decay. 
Another necessary condition for this  direct \CP violation to occur 
is interference of at least two amplitudes with different strong phases. 
The strong phase differences are known to be sizeable in charm decays and can enter through
the resonances that abundantly contribute to the four-body final states. 
The sensitivity to \CP violation is usually 
 best for decays where the strong phases between interfering resonances have large differences.
 A rich spectrum of resonances contributes to 
the decay \decay{\Dz}{\pip\pim\pip\pim}, which according to the amplitude analysis 
performed by the FOCUS collaboration~\cite{focus}, 
is dominated by the amplitudes for $\Dz \to a_1(1260)^+ \pi^-$ with $a_1(1260)^+ \to \rhoz(770) \pi^+$ and 
for $\Dz \to \rhoz(770) \rhoz(770)$. 

In the SM, violation of the \CP symmetry in the charm sector is expected at or below the $\mathcal{O}$($10^{-3}$) level~\cite{Grossman:2006jg}. 
Contributions from particles that are proposed to exist in extensions of the SM may participate in higher-order loop contributions (penguin diagrams)
and enhance the level of \CP violation. 
Multibody decays, such as  \decay{\Dz}{\pip\pim\pip\pim}, 
allow the \CP asymmetries to be probed across the phase space of the decay, 
and these local \CP asymmetries may be larger than global \CP asymmetries. 

The analysis of \decay{\Dz}{\pip\pim\pip\pim} is primarily sensitive to direct \CP violation. 
In addition to this direct \CP violation, the time-integrated
\CP asymmetry in \decay{\Dz}{\pip\pim\pip\pim} decays can also receive 
an indirect contribution 
arising from either  \Dz-\Dzb mixing or 
 interference between direct decays and decays following mixing. 
While direct \CP asymmetry depends on the decay mode, indirect \CP violation is expected 
to be  universal within the SM. 
The time-dependent measurements 
of \decay{\Dz}{\pim\pip},~\Km\Kp decays constrain the indirect \CP asymmetry 
below the $\mathcal{O}$($10^{-3}$) level~\cite{HFAG}, 
which is beyond the sensitivity of this analysis.

Previously, the most sensitive search for \CP violation in the \decay{\Dz}{\pip\pim\pip\pim} decay 
was performed by the \lhcb collaboration with data collected in 2011~\cite{LHCb-PAPER-2013-041}. 
A binned \chisq technique ($S_{\CP}$) was used to exclude \CP-violating effects at the 10\% level. 
Four-body decays require five independent variables to fully represent the phase space. 
Consequently, binned analyses will have a trade-off between minimising the number 
of bins, in order to maximise the number of events per bin, and retaining sensitivity to the interference between all contributing resonances. 
The measurement presented here includes data collected in 2012, resulting in a signal sample 
that is about three times larger than in the previous LHCb analysis. The method exploited here, known as the energy test, 
is an unbinned technique, which is advantageous in the analysis of multibody decays. 

The energy test was applied for the first time to search for \CP violation in decays of 
~\decay{\Dz}{\pim\pip\piz}~\cite{LHCb-PAPER-2014-054}; here we present its first application to four-body decays.
The energy test is used to assess the compatibility of the observed data with \CP symmetry.
It is sensitive to local \CP violation in the phase space and not to
global asymmetries, which may also arise from different production  cross-sections of \Dz and \Dzb mesons at a
proton-proton collider. The method identifies the phase space regions in which \CP violation is observed. 
Being model-independent, this method neither identifies which amplitudes 
contribute to the observed asymmetry nor measures the actual
asymmetry. Consequently,  a model-dependent amplitude analysis of \Dz and \Dzb decays would 
be required if evidence for a non-zero \CP asymmetry is
obtained. 

The analysis presented here probes separately
both $P$-even and $P$-odd \CP asymmetries. 
The $P$-even test is performed through the comparison of the distribution of events in the \Dz and \Dzb phase 
spaces, characterised by squared invariant masses. 
Additionally characterising the events using  a triple product of final-state 
particle momenta~\cite{Valencia,Datta:2003mj,EnergyTest2016} 
gives sensitivity to $P$-odd amplitudes, and thus 
allows the first test for $P$-odd \CP asymmetries in an unbinned 
model-independent technique.

\section{Detector and reconstruction}
\label{sec:Detector}

The \lhcb detector~\cite{Alves:2008zz,LHCb-DP-2014-002} is a single-arm forward
spectrometer covering the \mbox{pseudorapidity} range $2<\eta <5$,
designed for the study of particles containing \bquark or \cquark
quarks. The detector includes a high-precision tracking system
consisting of a silicon-strip vertex detector surrounding the $pp$
interaction region, a large-area silicon-strip detector located
upstream of a dipole magnet with a bending power of about
$4{\mathrm{\,Tm}}$, and three stations of silicon-strip detectors and straw
drift tubes placed downstream of the magnet.
The tracking system provides a measurement of momentum, \ptot, of charged particles with
a relative uncertainty that varies from 0.5\% at low momentum to 1.0\% at 200\gevc.
The minimum distance of a track to a primary $pp$ interaction vertex (PV), 
the impact parameter (IP), is measured with a resolution of $(15+29/\pt)\mum$,
where \pt is the component of the momentum transverse to the beam, in\,\gevc.
Different types of charged hadrons are distinguished using information
from two ring-imaging Cherenkov detectors. 
Photons, electrons and hadrons are identified by a calorimeter system consisting of
scintillating-pad and preshower detectors, an electromagnetic
calorimeter and a hadronic calorimeter. Muons are identified by a
system composed of alternating layers of iron and multiwire
proportional chambers. 

This analysis uses the data from $pp$ collisions collected by the LHCb experiment in 2011 and 2012 corresponding 
to integrated luminosities of, respectively, $1\invfb$ and $2\invfb$  at centre-of-mass energies of $7\tev$ and $8\tev$. 
The polarity of the dipole magnet is reversed periodically throughout data-taking.
The configuration with the magnetic field pointing upwards (downwards), bends positively (negatively)
charged particles in the horizontal plane towards the centre of the
LHC. Similar amounts of data were recorded with each polarity,
which reduces the 
effect of charge-dependent detection and reconstruction
efficiencies on results obtained from the full data sample.

In the simulation, $pp$ collisions are
generated using \pythia 8~\cite{Sjostrand:2007gs} with a specific \lhcb
configuration~\cite{LHCb-PROC-2010-056}. Decays of hadronic particles are described by
\evtgen~\cite{Lange:2001uf}. 
 The interaction of the generated particles with the detector
and its response are implemented using the \geant toolkit~\cite{Allison:2006ve, *Agostinelli:2002hh}
as described in Ref.~\cite{LHCb-PROC-2011-006}.

The online event selection is performed by a trigger, 
 which consists of a hardware stage, based on high-\pt signatures  
from the calorimeter and muon systems, followed by a two-level
software stage.
Events are required to pass both hardware and software trigger levels.
The software trigger at its first level applies partial event reconstruction. 
It requires at least one good quality track associated with a particle having high \pt 
and not originating from a PV.

A second-level software trigger, optimised for four-body hadronic charm decays, fully reconstructs 
\decay{\Dz}{\pip\pim\pip\pim} candidates coming from $\decay{\Dstarp}{\Dz\pi_s^+}$ decays. 
The charge of the soft pion ($\pi_s$)  tags the flavour of the \D mesons at production, which is needed as \pip\pim\pip\pim is 
a self-conjugate final state accessible to both \Dz and \Dzb decays. 
The trigger selection ensures the suppression of combinatorial
background while minimising the distortion of the acceptance in the
phase space of the decay. The trigger requires a four-track secondary
vertex with all tracks being of good quality and passing minimum
momentum and transverse momentum requirements. The pions from 
the candidate \Dz decay are required to have a 
large impact parameter significance ($\chi^2_{\rm IP}$)  with respect to all PVs, 
where  $\chi^2_{\rm IP}$ is defined as the difference in the vertex-fit $\chi^2$ of  a 
PV reconstructed with and without the considered track. 
A part of the data collected in 2011 was taken with a different second-level trigger selection. In 
this selection only \decay{\Dz}{\pip\pim\pip\pim} candidates are reconstructed while the 
\Dstarp reconstruction is performed only during the offline
selection. For a part of the data collected in 2012, additional
events were selected  in the trigger from the partial reconstruction of \decay{\Dz}{\pip\pim\pip\pim} candidates 
arising from $\decay{\Dstarp}{\Dz\pi_s^+}$ decays, using only information from  one \pip\pim pair
and a soft pion.

\section{Offline event selection}
\label{sec:Selection}

In the offline selection, 
signal candidates are required to be associated to candidates that passed the online selection described in the previous section.
In addition, the offline selection imposes more stringent 
kinematic criteria than those applied in the trigger. The \Dstarp and \Dz 
candidates must have $\pt>500\mevc$.  All the candidate pion tracks,
those from the \Dz decay products and the $\pi_s$ mesons, are required to have
$\pt>350\mevc$ and $\ptot>3\gevc$  to reduce the combinatorial background. 
The candidate \Dz decay products must form a good quality vertex. 
As a consequence of the non-negligible \Dz lifetime, the \Dz decay 
vertex should typically be significantly displaced from the PV; this
is ensured by applying a selection on the significance of the \Dz
candidate flight distance. Charm mesons from \bquark hadron decays 
have larger IPs due to the comparatively long $b$ hadron lifetimes.  
This secondary charm contribution is suppressed by imposing an upper
limit on the $\chi^2_{\rm IP}$  of the \Dz candidate.  Background
from \decay{\Dz}{\Km\pip\pim\pip} decays, with a kaon misidentified as a pion, 
 is reduced by placing tight requirements on the \pipm particle identification 
based on the ring-imaging Cherenkov detectors. The contribution of the
Cabibbo-favoured 
\decay{\Dz}{\KS\pip\pim} decay is found to be below the percent
level. As investigated in Ref.~\cite{LHCb-PAPER-2013-041}, partially reconstructed or misreconstructed 
multibody $\D_{(s)}$ decays (\eg,\ decays with a missing pion or a kaon misidentified as pion) 
do not give rise to peaking backgrounds under the
\decay{\Dz}{\pip\pim\pip\pim} signal.

Constraints on the decay kinematics are applied to improve mass and
momentum resolutions. The four pions from the \Dz decay are constrained to come from a
common vertex and the decay vertex of the \Dstarp candidate is constrained to coincide
with its PV~\cite{Hulsbergen:2005pu}. These constraints are applied in the determination of the
mass difference, \dm,  between the \Dz and the \Dstarp. The
\Dz is constrained to its nominal mass in the determination of the  kinematics in the
\decay{\Dz}{\pip\pim\pip\pim} decay. A requirement on fit quality for the \Dstarp vertex
fits efficiently suppresses combinatorial background.  This requirement also suppresses 
the contribution from \Dstarp mesons originating 
from long-lived \bquark hadrons. The remaining component in the analysis from this
source is not sensitive to \CP asymmetries in \bquark hadrons as
the flavour tag is obtained from the charge of the $\pi_s$ in the decay of the \Dstarp meson.

The $\pi_{s}$ is a low-momentum particle, with the consequence that the large deflection
 in the magnetic field leads to different acceptances for the two charges.
Consequently, the soft pion is restricted to the
region where the detection asymmetry is small. This is achieved through the application of
fiducial cuts on the soft pion momentum, following Ref.~\cite{Aaij:2011in}.
As the kinematics of the slow pion are largely uncorrelated with the \Dz phase space, 
the $\pi_{s}$ detection asymmetry would result in a global asymmetry to which this analysis is not sensitive. 
There are, however, differences in the detection efficiencies of the \Dz and \Dzb daughters that may introduce 
additional asymmetries localised in the phase space of \Dz decays, 
and which are discussed in detail in Sect.~\ref{sec:Systematics}.

\begin{figure}[tb!]
  \centering
  \includegraphics[height=0.46\textwidth]{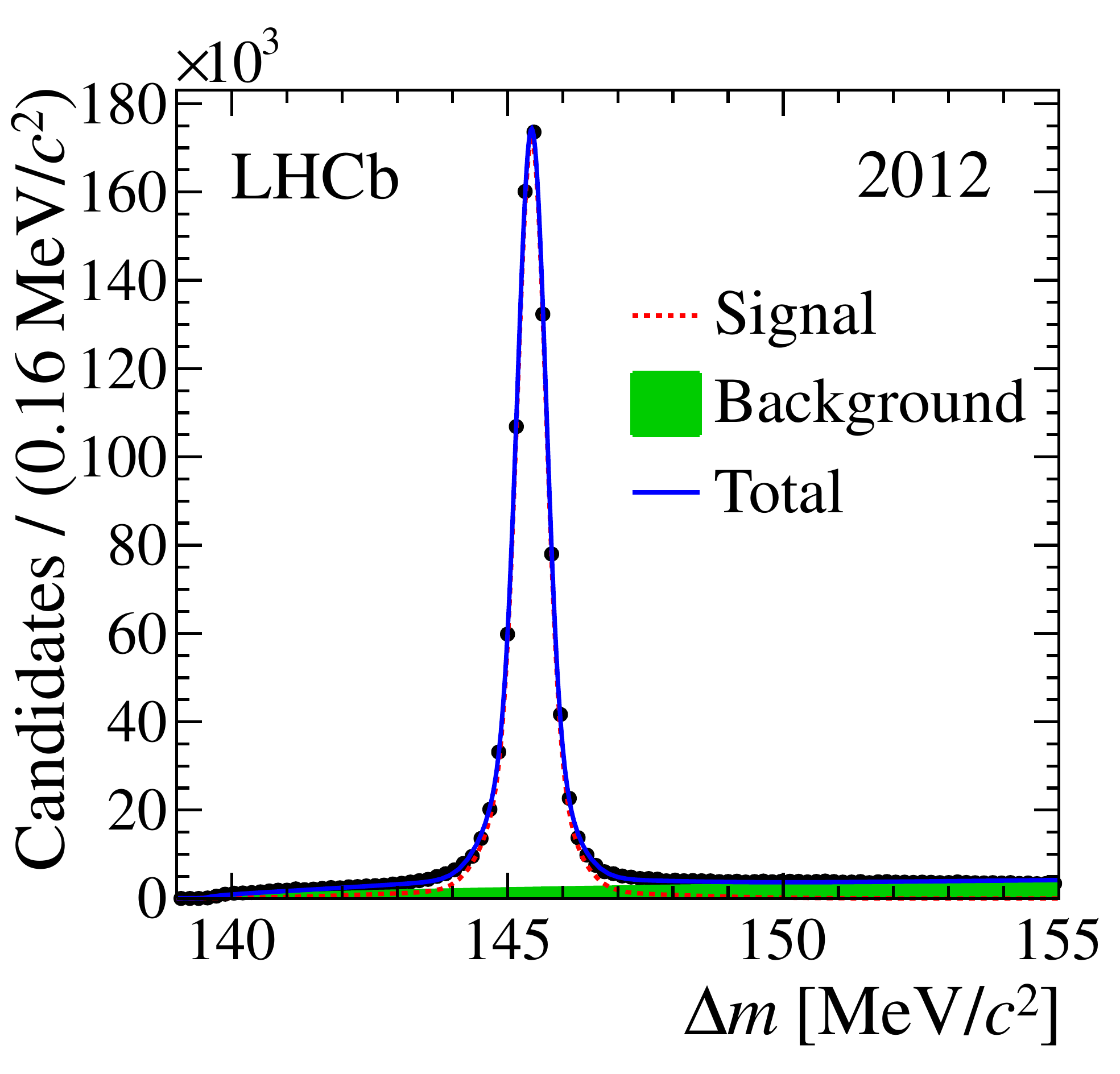}
  \caption{\small Distribution of \dm with fit overlaid for the selected \Dstarp candidates 
  in the 2012 data. The data points and the contributions from signal, 
background, and their total obtained from the fit are shown. 
\label{fig:deltam}}
\end{figure}

The signal region in the \Dz invariant mass is defined as 
$1852<m(\pip\pim\pip\pim)<1882\mevcc$, corresponding to a full range of about four times the mass resolution.
The signal yield is estimated from the \dm distribution, which is shown in 
Fig.~\ref{fig:deltam} for the 2012 data. 
These \dm distributions are modelled by the sum of three 
Gaussian functions for signal and a 
second-order polynomial multiplied by a threshold function $\sqrt{1-m_{\pi}/\dm}$, where $m_{\pi}$ is the pion mass,  
describing combinatorial and random soft-pion backgrounds. 
The selected 
samples comprise 320,000 and 720,000 signal candidates in 
the 2011 and 2012 data with purities of $97\%$ and $96\%$, respectively.
The final signal sample is selected requiring $|\dm -145.44| <0.45 \mevcc$, which corresponds 
to selecting a region with a width roughly twice the effective \dm resolution. 

\section{Description of the phase space}
\label{sec:Coordinates}

Five coordinates are required for a full description of the phase
space of four-body decays~\cite{Byckling}.  In contrast to three-body decays, there is no standard or commonly
preferred choice of coordinates. Two-body and three-body invariant
mass combinations of the pions are used as coordinates here. The energy test performed here is a statistical method comparing 
 the distributions of \Dz and \Dzb candidates in phase space (see Sect.~\ref{sec:Method}). 
Therefore, it is sensitive to the position of an event 
in phase space and to the choice of coordinates spanning this phase
space.
The choice of coordinates influences the sensitivity of the analysis 
as it will change the distance between events in the phase space. 
Furthermore, \decay{\Dz}{\pip\pim\pip\pim} decays contain two \pip and two \pim mesons. The pions of the same charge can be interchanged; as a result each decay 
can be placed at four points in phase space. The energy test is
sensitive to such pion interchange. 
To obtain both a unique output and optimal sensitivity from the energy test, 
 an ordering of
the input variables of the energy test is defined in the following.

The order of the charges of the four pions in the \Dz decay $\pi_1\pi_2\pi_3\pi_4$ is fixed to 
\pip\pim\pip\pim.\footnote{In a \Dzb decay the order of
  $\pi_1\pi_2\pi_3\pi_4$ is charge-conjugated, \pim\pip\pim\pip,
  with respect to that of the \Dz decay.}
 There are four two-body combinations in which resonances can be formed and these are the \pip\pim pairs: 
$\pi_1\pi_2$, $\pi_1\pi_4$, $\pi_3\pi_2$, and $\pi_3\pi_4$.  Likewise, there are four 
three-body combinations: the two of positive charge, $\pi_1\pi_2\pi_3$ and $\pi_1\pi_3\pi_4$, 
and the two of negative charge, $\pi_1\pi_2\pi_4$ and $\pi_2\pi_3\pi_4$. 
The invariant masses of all possible \pip\pim pairs are calculated and sorted for each event.
The \pip\pim pair with the largest invariant mass is fixed to be $\pi_3\pi_4$, 
which fully determines the order of all four pions. 
As only a small fraction of the $\rho(770)$ resonance, either produced directly from the \Dz decay or 
through $a_1(1260)$ decays, contributes to the largest $m(\pip\pim)$,
the $\pi_3\pi_4$ combination is excluded from the coordinates used. 
While any combination of five variables covers the full phase space, 
 the choice made here is to keep variables sensitive to the presence of the main resonances.
The two-body and three-body mass combinations 
that do not contain the $\pi_3\pi_4$ pair are used
for the energy test coordinates.  This results in five invariant masses, 
$\pi_1\pi_2$, $\pi_1\pi_4$, $\pi_2\pi_3$, $\pi_1\pi_2\pi_3$ and $\pi_1\pi_2\pi_4$,
which are expected to cover most of the intermediate resonance contributions. 

The choice of using only invariant masses has a limitation. Invariant
masses are even under parity transformation, and a comparison of
the \Dz and \Dzb samples probes only $P$-even \CP asymmetries. In four-body decays, however,
$P$-odd amplitudes can also be present.
In \decay{\Dz}{\pip\pim\pip\pim} decays, there is only one significant $P$-odd amplitude,
the perpendicular helicity ($A_{\perp}$) of $D^0\to\rhoz(770) \rhoz(770)$ decays.
Alternatively, in the partial-wave basis, it is the amplitude corresponding to 
the P-wave $D^0\to\rhoz(770) \rhoz(770)$ decays. Its contribution to the total \decay{\Dz}{\pip\pim\pip\pim} width 
is about 6\%~\cite{PDG2016}. The default approach is extended to make a complementary 
test of the $P$-odd \CP asymmetry,
which may arise from interference between $P$-odd and $P$-even
amplitudes. This is discussed further in Ref.~\cite{EnergyTest2016}.

Triple products,  which are by definition $P$-odd, can be used to probe $P$-odd \CP violation~\cite{Aaij:2014qwa}.
These asymmetries are proportional to the cosine  
of the strong-phase difference between the interfering partial waves~\cite{Valencia}, and thus will be enhanced where $P$-even \CP asymmetries, 
proportional to the sine of the strong-phase difference, lack sensitivity.
A triple product $C_T=\vec{p}_1 \cdot (\vec{p}_2 \times \vec{p}_3)$ is constructed for \Dz decays, where pion momenta are measured in the \Dz rest frame.
 Here $\vec{p}_1$, $\vec{p}_2$ and $\vec{p}_3$ are the vector momenta of $\pi_1$, $\pi_2$ and $\pi_3$ sorted as described above 
(\ie,\ $\pi_4$ is excluded). 
The corresponding triple product for the \Dzb decays is obtained by
applying the \CP  transformation, 
$\CP(C_T)=-C(C_T)=-\overline{C}_T$. The $\overline{C}_T$ observable is constructed
 by charge conjugating the pions entering $C_T$
(\ie, the excluded pion in $\overline{C}_T$ is the \pip in the largest $m(\pip\pim)$ combination).
The total sample is divided into four subsamples according to the \Dz flavour and the sign of the triple product:
\begin{equation}
\label{eqn:tp_split}
[\rm{I}] \ \textit{\Dz}(\textit{C}_\textit{T}>0), \ \ [\rm{II}] \ \textit{\Dz}(\textit{C}_\textit{T}<0), \ \ [\rm{III}] \ \textit{\Dzb}(-\overline{\textit{C}}_\textit{T}>0), \ \ [\rm{IV}] \ \textit{\Dzb}(-\overline{\textit{C}}_\textit{T}<0).
\end{equation}

Samples $\rm{I}$ and $\rm{III}$ are related by the \CP transformation, and so are $\rm{II}$ and $\rm{IV}$.
The yields of these four samples are listed in Table~\ref{tab:yields}.
The test for the presence of $P$-even \CP asymmetry is performed by comparing  the
combined sample $\rm{I}+\rm{II}$ with the combined sample $\rm{III}+\rm{IV}$. This
corresponds to the integration over $C_T$ and is the default test, in
which \Dz and \Dzb samples are compared in the phase space spanned with
invariant masses only. Similarly, the test for a $P$-odd \CP asymmetry
is performed by comparing the combined sample $\rm{I}+\rm{IV}$ with the
combined sample $\rm{II}+\rm{III}$.
This comparison is performed in the same phase space as the default
$P$-even approach and allows the $P$-odd contribution to the
\CP asymmetry to be probed,  since the $P$-even contribution cancels~\cite{EnergyTest2016}.
No triple-product asymmetry
measurements exist for \decay{\Dz}{\pip\pim\pip\pim} decays and the
previous \lhcb study~\cite{LHCb-PAPER-2013-041} was performed in the phase space based on the
invariant masses only. Consequently, this is the first time a $P$-odd
\CP asymmetry is  investigated in this decay mode.

\begin{table}[t!]

\caption{The yields of signal events in the four samples that obtained from fits to the $\Delta m$ distribution.}
\centering

\begin{tabular}{c|cccc}
\hline
Sample & $\rm{I}$ & $\rm{II}$ & $\rm{III}$ &$\rm{IV}$\\
\hline  
Yields & 256\,466$\pm$629 & 246\,629$\pm$519 & 258\,274$\pm$574 & 246\,986$\pm$607\\
\hline  
\end{tabular}

\label{tab:yields}
\end{table}

\section{Energy test}
\label{sec:Method}

Model-independent searches for local \CP violation are typically
carried out using a binned \chisq  approach to compare the relative
density in a bin of phase space of a decay with that of its \CP-conjugate.
This method was used in a previous study of  \decay{\Dz}{\pip\pim\pip\pim}  decays~\cite{LHCb-PAPER-2013-041}. 
As discussed in the previous section, five
coordinates are required to describe four-body decays. 
A model-independent unbinned statistical
method called the energy test was introduced in
Refs.~\cite{doi:10.1080/00949650410001661440,Aslan2005626}. The
potential for increased sensitivity of this method over binned \chisq
analyses in Dalitz plot analyses was shown in Refs.~\cite{Williams:2011cd,EnergyTest2016} 
and it was first applied to experimental data in Ref.~{\cite{LHCb-PAPER-2014-054}}.

This Letter introduces the first application of the energy test technique to four-body decays, 
where it is used to compare two event samples in tests of both $P$-even and $P$-odd type \CP violation.
The $P$-even energy test separates events 
according to their flavour, and then compares these \Dz and \Dzb samples. The $P$-odd
energy test separates events using both their flavour
and sign of the triple product, as described in the previous section.

A test statistic, $T$, is used to compare the
average distances of events in phase space. The variable $T$ is
based on a  function 
$\psi_{ij} \equiv \psi(d_{ij})$ which depends on the distance $d_{ij}$ between events $i$ and $j$.
It is defined as
\begin{equation}
    T = \sum_{i,j>i}^{n}\dfrac{\psi_{ij}}{n(n-1)}
    + \sum_{i,j>i}^{\overline{n}}\dfrac{\psi_{ij}}{\overline{n}(\overline{n}-1)}
    - \sum_{i,j}^{n,\overline{n}}\dfrac{\psi_{ij}}{n\overline{n}} ,
    \label{eqn:T}
\end{equation}
where the first and second terms correspond to an average weighted distance
between events within the $n$ events of the first sample and between the $\overline{n}$
events of the second sample, respectively.
The third term measures the average weighted distance between events in the first sample and events
in the second sample. 
If the distributions of events in both samples are identical, $T$
will randomly fluctuate around a value close to zero.

The normalisation factors in the denominators of the terms of Eq.~\ref{eqn:T} remove the impact of global asymmetries 
between \Dz and \Dzb samples. In the $P$-odd test, subsamples of both \Dz and \Dzb samples are 
combined. Consequently, any global asymmetries in these could result in
local asymmetries in the samples used for the $P$-odd test. 
Therefore, for the $P$-odd test the global asymmetry between \Dz and \Dzb is removed by randomly rejecting
some of the \Dzb candidates to equalise the size of the samples of the two flavours before combining the events into the two samples that are compared.

The function $\psi$ should decrease with increasing distance $d_{ij}$ between events $i$ and $j$,
in order to increase the sensitivity to local asymmetries. A Gaussian function is chosen,
$\psi(d_{ij})=e^{-d_{ij}^2/2\delta^2}$, with a
tuneable parameter $\delta$ (see Sect.~\ref{sec:Sensitivity}) that
describes the effective radius in phase space within which
a local asymmetry is measured. Thus, this parameter should be larger than the resolution of $d_{ij}$
but small enough not to dilute locally varying asymmetries. The distance between two points 
is obtained using the five squared invariant masses discussed in the previous section 
and calculated as 
\begin{equation}
    d_{ij}^2=(m_{12}^{2,j}-m_{12}^{2,i})^2+(m_{14}^{2,j}-m_{14}^{2,i})^2+(m_{23}^{2,j}-m_{23}^{2,i})^2+
    (m_{123}^{2,j}-m_{123}^{2,i})^2+(m_{124}^{2,j}-m_{124}^{2,i})^2.
\end{equation}

In the case of \CP violation, the average distances entering in the third term
of Eq.~\ref{eqn:T} are larger than in the other terms. Due to the characteristics of the $\psi$ function,
this leads to a reduced magnitude of this third term relative to
the other terms. Therefore, larger \CP asymmetries lead to larger values of $T$.
This is translated into a \pvalue under the hypothesis of \CP symmetry by comparing
the  $T$ value observed in data to a distribution of $T$ values obtained from permutation
samples. The permutation samples are constructed by randomly
assigning events to either of the samples, thus simulating a situation without \CP violation.
The \pvalue for the no-\CP-violation hypothesis is obtained as the fraction of permutation $T$ values greater
than the observed $T$ value.

For scenarios where the observed $T$ value lies well within the range of permutation $T$ values, the \pvalue can be calculated by simply counting how many permutation $T$ values are larger than the observed one. 
If large \CP violation is observed, the observed $T$ value is likely to lie outside
the range of permutation $T$ values. In this case the permutation $T$ distribution can be fitted
with a generalised-extreme-value (GEV) function, as demonstrated in
Refs.~\cite{doi:10.1080/00949650410001661440,Aslan2005626} and
used in Ref.~\cite{LHCb-PAPER-2014-054}.
The \pvalue from the fitted $T$ distribution can be calculated as the fraction of the integral
of the function above the observed $T$ value.
The uncertainty on the \pvalue is obtained by randomly resampling the fit parameters
within their uncertainties, taking into account their correlations, and by extracting a \pvalue for each of
these generated $T$ distributions. The spread of the resulting \pvalue distribution is used to set 68\% confidence intervals. A 90\% confidence-level upper limit is quoted where no significantly non-zero \pvalue can be obtained
from the fit.

\begin{figure*}[bth!]
    \centering
    \includegraphics[width=0.39\textwidth]{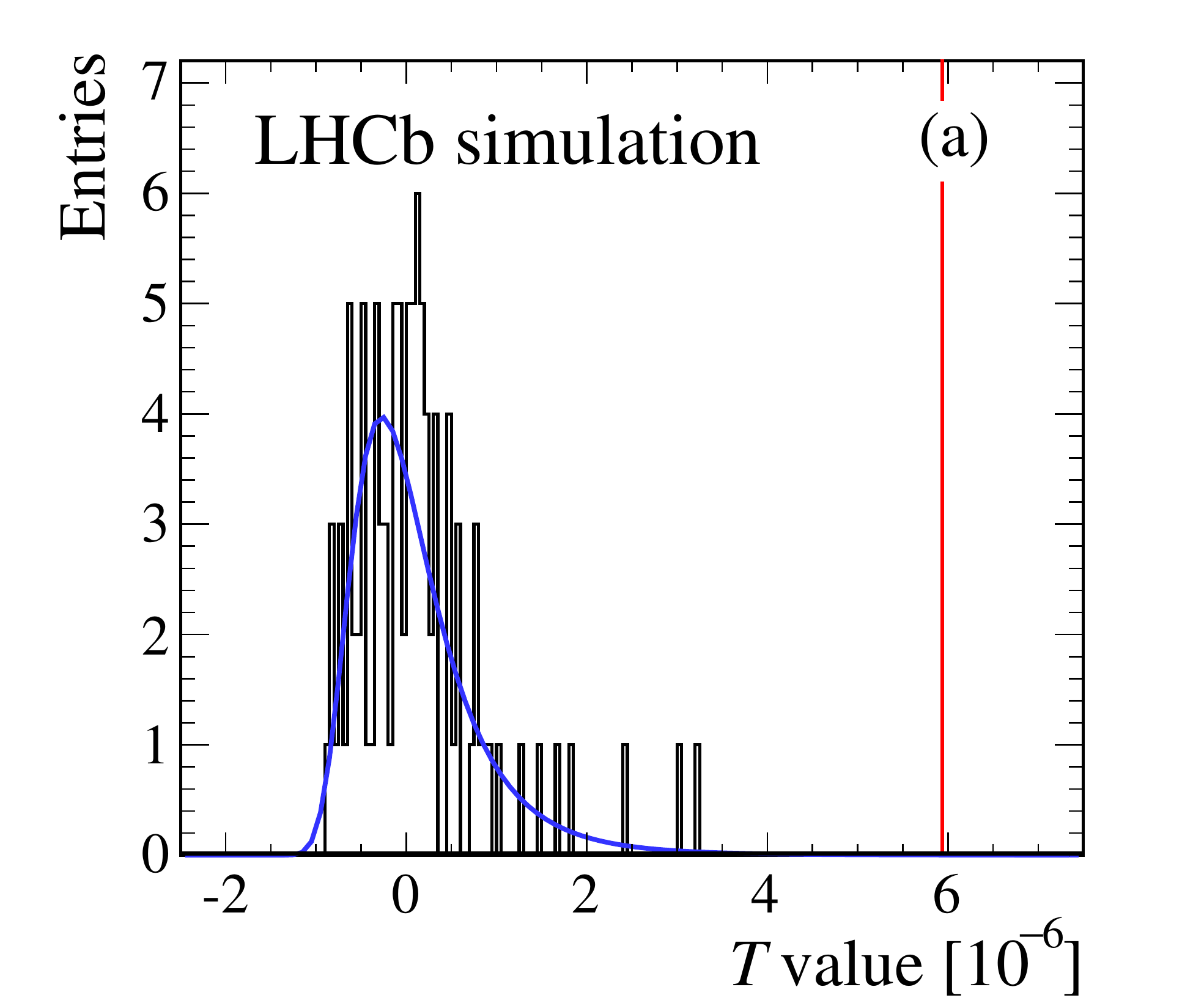}
    \hspace{1cm}
    \includegraphics[width=0.39\textwidth]{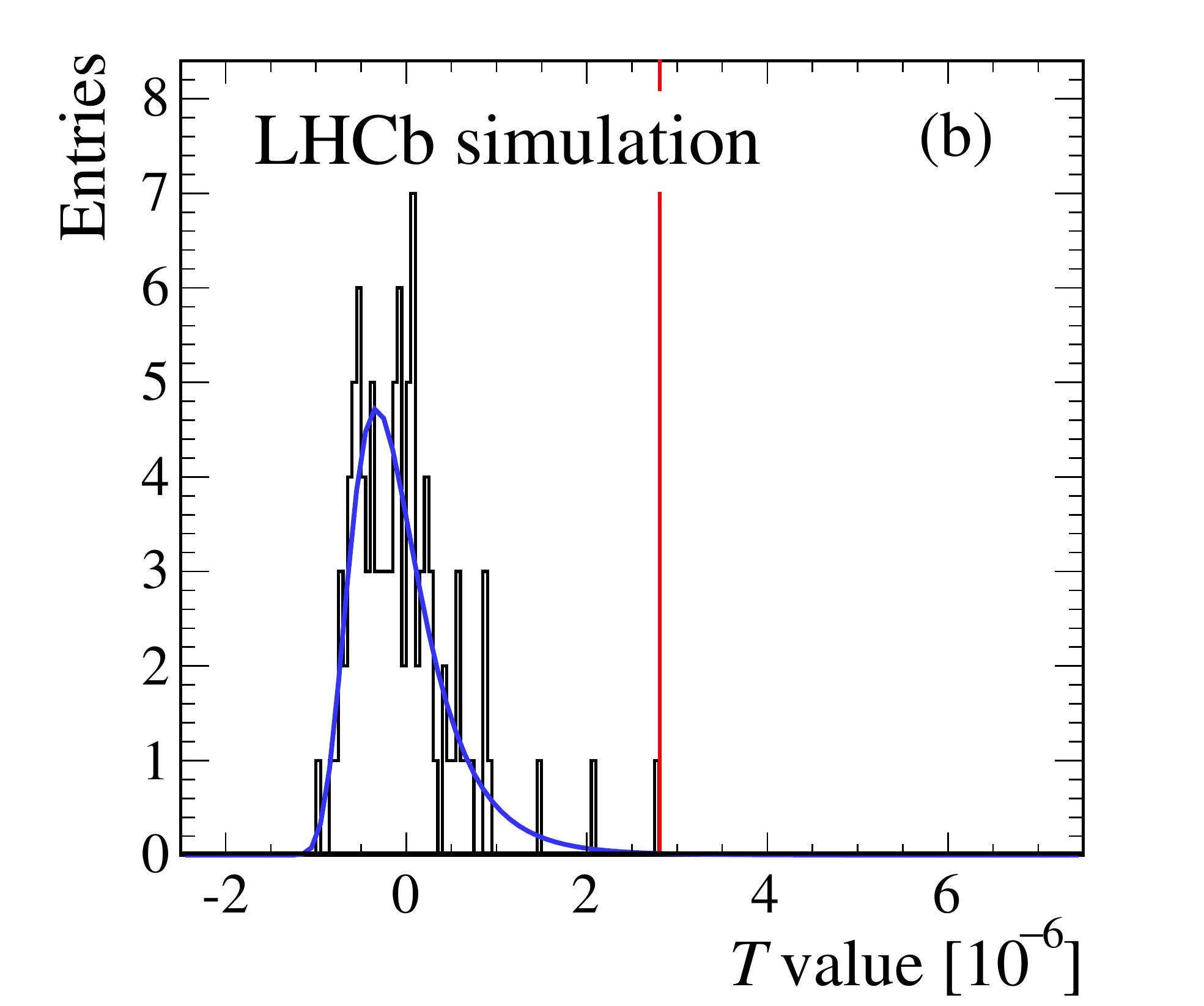}

    \includegraphics[width=0.39\textwidth]{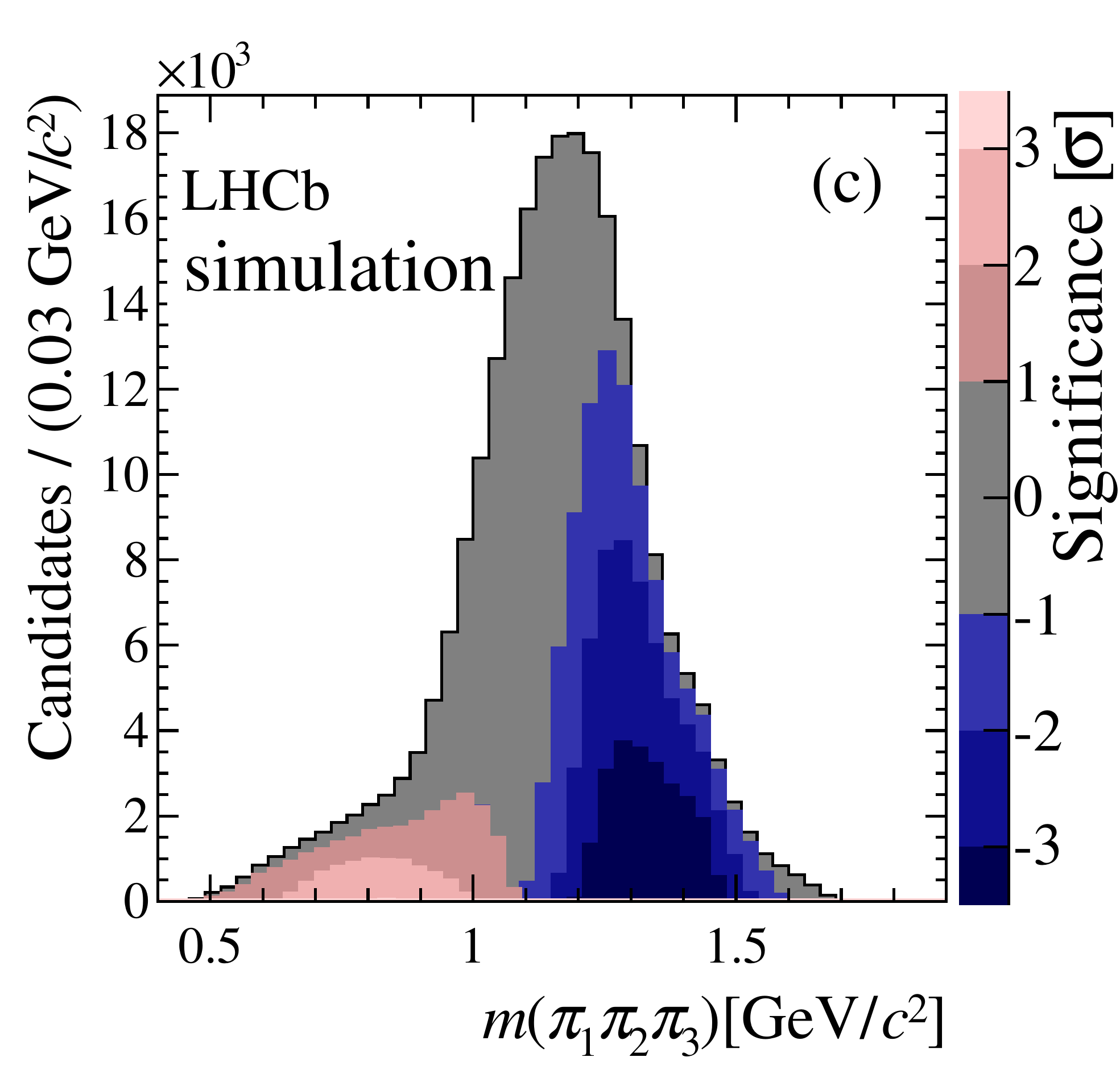}
    \hspace{1cm}
    \includegraphics[width=0.39\textwidth]{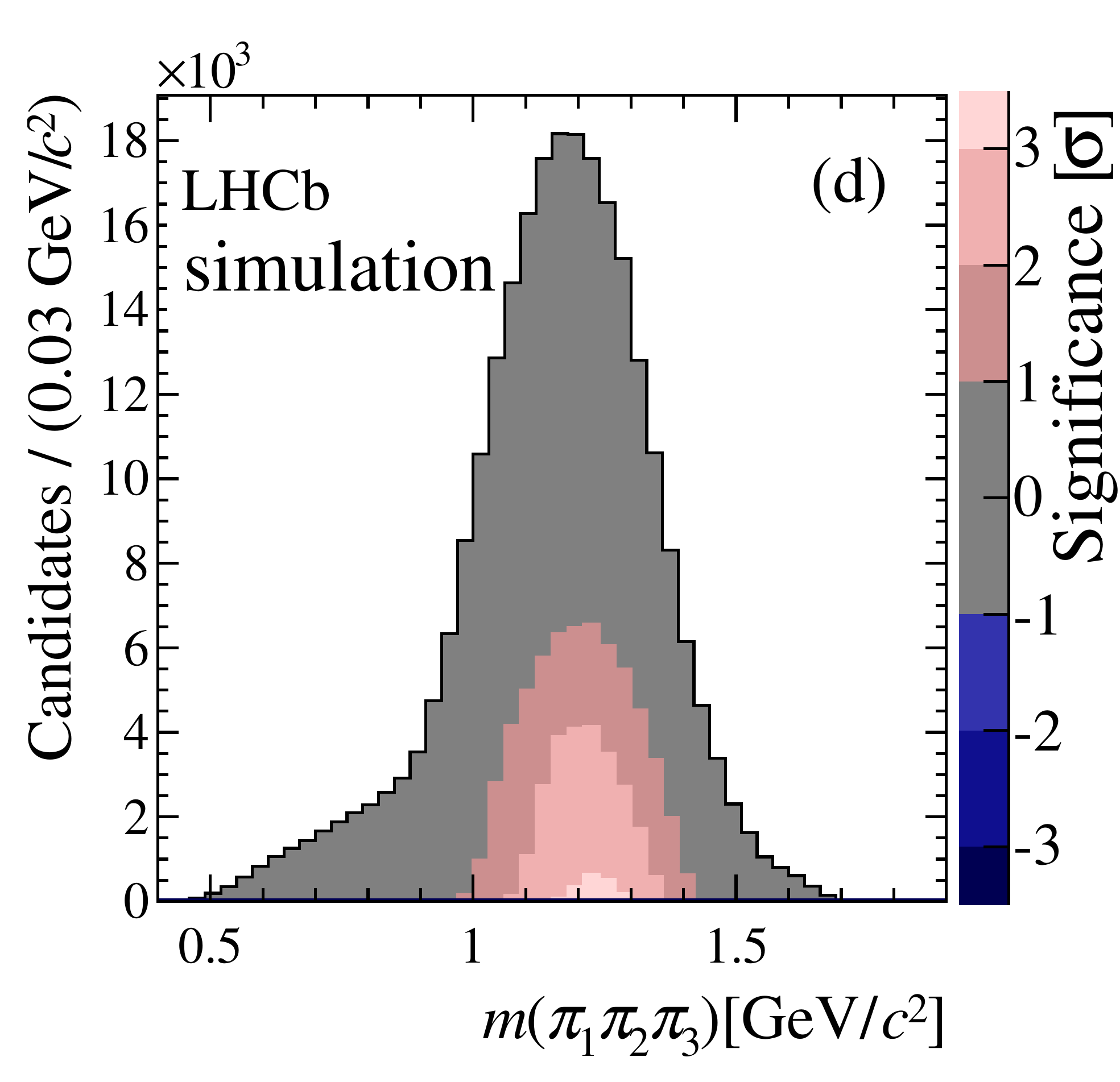}

    \includegraphics[width=0.39\textwidth]{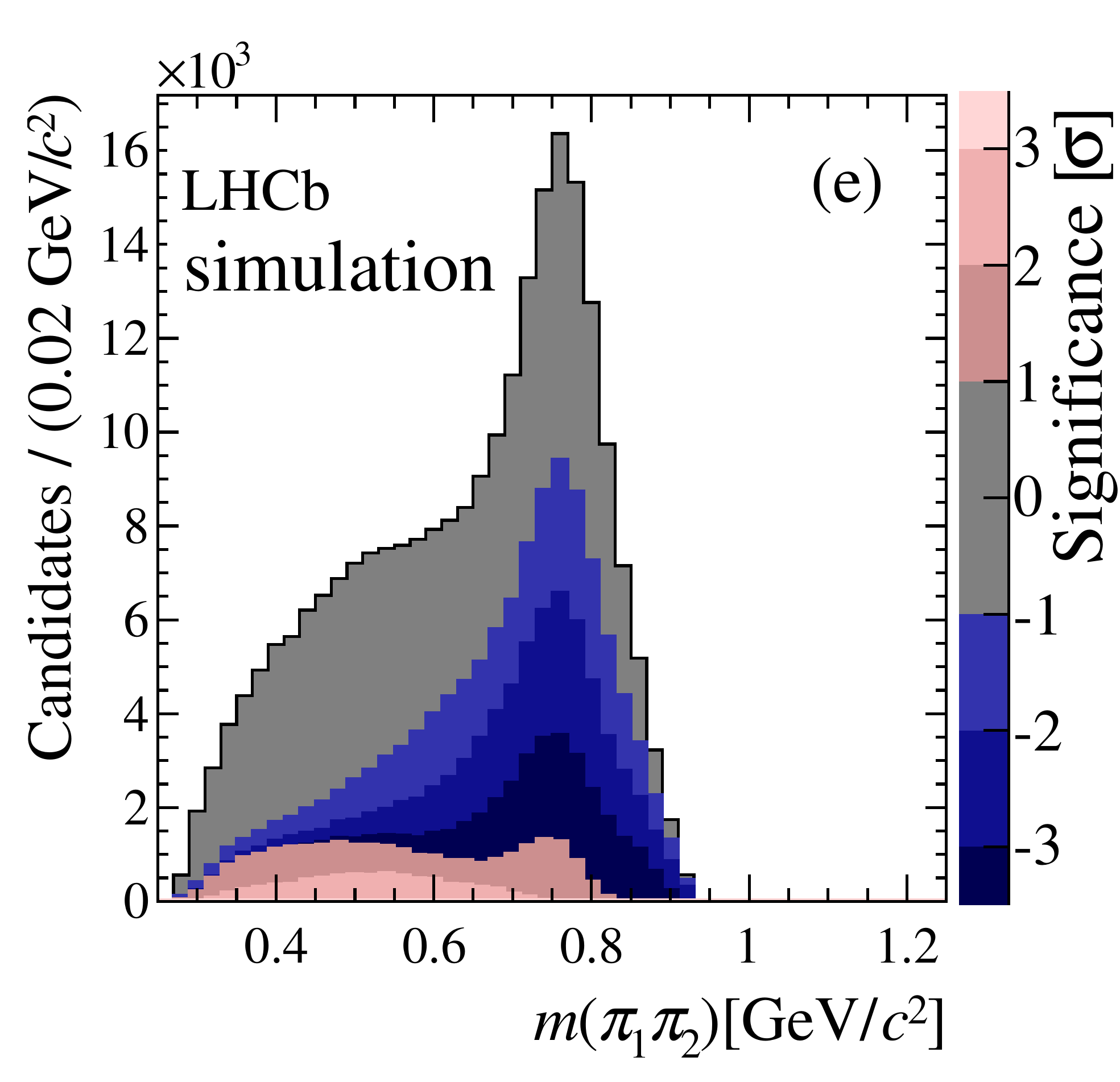}   
    \hspace{1cm}
    \includegraphics[width=0.39\textwidth]{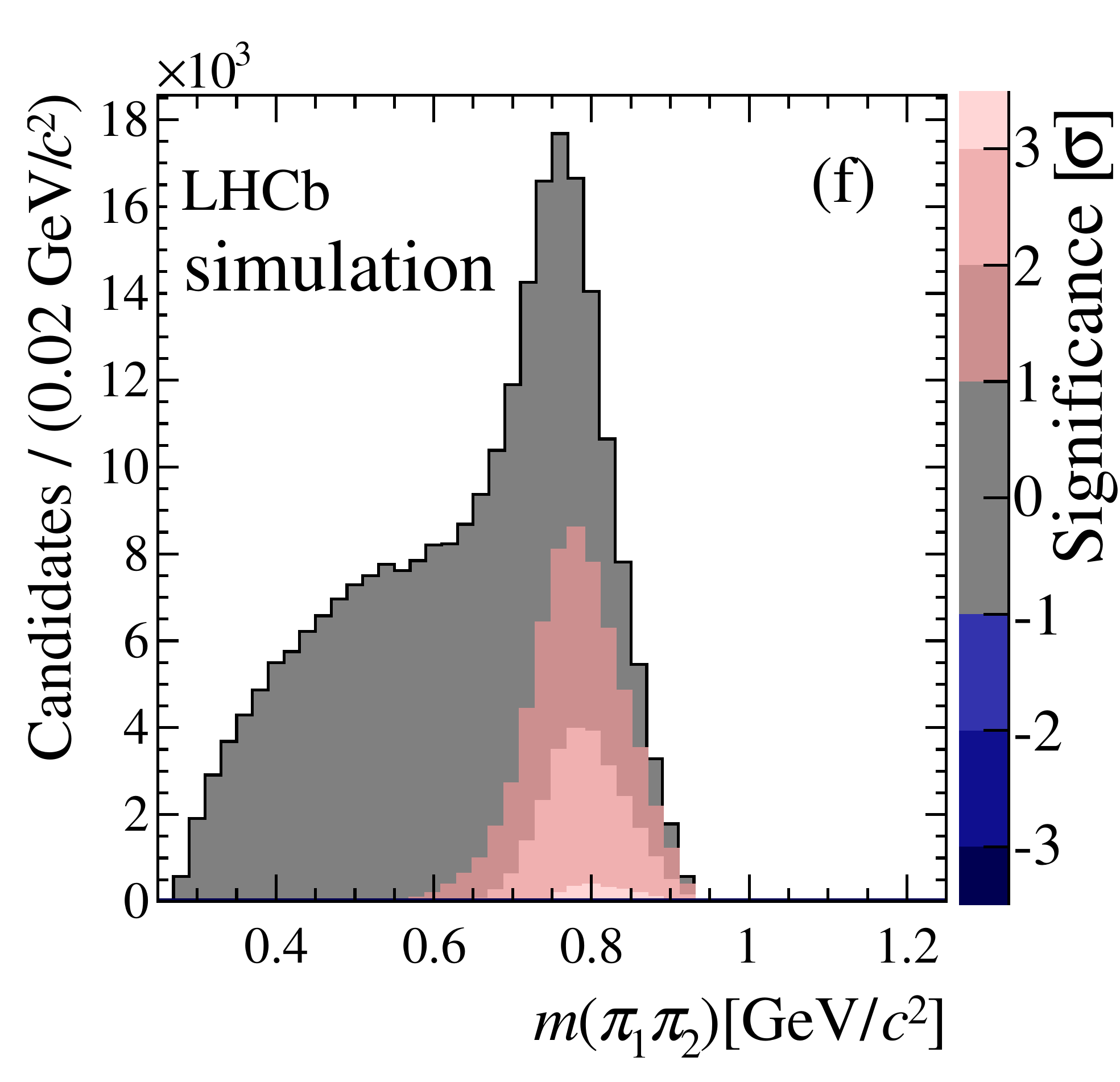}

    \caption{
        \small
        (a,b) Distribution of permutation $T$-values fitted with a GEV function for the simulated sample and showing
        the measured $T$-value as a vertical line, and (c,d,e,f) local asymmetry
        significances. Left column plots are for a $P$-even \CP-violation test
        with a $3^\circ$ phase \CP violation introduced in the $a_1(1260)^+$
        resonance (see text), projected onto the (c) $m(\pi_1 \pi_2 \pi_3)$
        and (e) $m(\pi_1\pi_2)$ axes. Right column plots are for a $P$-odd \CP-violation test with 
        $3^\circ$ phase \CP violation introduced in the P-wave $\rhoz(770)
        \rhoz(770)$ resonance projected onto the same axes. In plots
        (c,d,e,f)  the grey area corresponds to candidates with a
        contribution to the $T$-value of less than one~standard deviation.  The pink (blue) area corresponds to candidates with a
        positive (negative) contribution to the $T$-value. Light, medium
        or dark shades of pink and blue correspond to 
        between one and two, two and three, and more than
    three standard-deviation contributions, respectively.}
    \label{fig:method_visualisation}
\end{figure*}

The number of permutations is constrained by the available computing time.
The default \pvalue extraction, defined before obtaining the result
from the data, uses the counting method as long as at least three permutation $T$ values are found to be
larger than the observed $T$ value. Otherwise, the \pvalue is determined by integrating the fitted GEV function.
The \pvalues presented here are based on over 1000 permutations for the default data results and on 100 permutations
for the sensitivity studies (see Sect.~\ref{sec:Sensitivity}).

A visualisation of regions of significant asymmetry is obtained by assigning an asymmetry significance
to each event. The contributions of a single event in one sample,
$T_i$, and a single event in 
the other sample, $\overline{T}_i$, to the total $T$ value are given by

\begin{equation}
    \label{eq:Ti}
    T_{i}  =  \dfrac{1}{2n\left( n-1\right) }\sum_{j\neq i}^{n}\psi_{ij}-\dfrac{1}{2n\overline{n} }\sum_{j}^{\overline{n}}\psi_{ij},
\end{equation} 
\begin{equation}
    \overline{T}_i  =  \dfrac{1}{2\overline{n}\left( \overline{n}-1\right) }\sum_{j\neq i}^{\overline{n}}\psi_{ij}-\dfrac{1}{2n\overline{n} }\sum_{j}^{n}\psi_{ij}.  
\end{equation}

Having obtained the $T_i$ and $\overline{T}_i$ values for all events, the permutation method is also used here to
define the significance of each event. 
For a given event $i$ the expected $T_i$ distribution in the case of
\CP symmetry is obtained by using the permutation method. The
distributions of the smallest negative and largest positive $T_i$
values of each permutation, $T_{i}^{\rm min}$ and $T_{i}^{\rm max}$,
are used to assign significances of negative and positive asymmetries,
respectively. If for real data the $T_i$ value falls in the right
(left) tail of the distribution of   $T_{i}^{\rm max}$ ($T_{i}^{\rm
min}$)  containing 32\%, 5\% or 0.3\% of the values it is assigned a
positive (negative) significance of 1, 2 or 3~$\sigma$, respectively.
The same procedure  
is applied to the $\overline{T}_i$ distribution, leading to a
phase space with an inverted asymmetry pattern. This method is illustrated in Sect.~\ref{sec:Sensitivity}.

The practical limitation of this method is that the number of mathematical operations scales quadratically
with the sample size. Furthermore, a significant number of permutations is required to get a sufficient
precision on the \pvalue. In this analysis, the method is implemented using parallel computing on graphics
processing units (GPUs)~\cite{Thrust}.

\section{Sensitivity studies}
\label{sec:Sensitivity}

To interpret  the results,  a study
of the sensitivity of the present data sample to different types of \CP violation is required.
The sensitivity is examined based on simplified simulation samples generated according to 
a preliminary version of the model in Ref.~\cite{dArgent:2017rbp}
based on CLEO-c measurements. 
The generation is
performed using MINT,  a software package for amplitude analysis of
multibody decays that has also been used by the CLEO
collaboration~\cite{Artuso:2012df}. Given the high purity obtained
in the selection, backgrounds are neglected in these
sensitivity studies. 

The variation of the selection efficiency  across phase space is taken
into account in these studies. 
This efficiency is measured using a sample of events
based on the full \lhcb detector simulation. 
The efficiency varies  mainly as a function of the momentum of the
lowest momentum pion in the event in the \Dz rest frame, and this efficiency variation is parameterised.
However, the dependence of the efficiency on this parametrisation is relatively weak.
For further studies the efficiency, based on the parametrisation, is
then applied to the simplified simulated data sets.

\renewcommand{\arraystretch}{1.1}
\begin{table}[b!]
    \caption{Overview of sensitivities to various \CP-violation scenarios
        in simulation.
        $\Delta A$ and $\Delta \phi$ denote, respectively,
        the  relative change in magnitude and  the change in phase of the amplitude of the resonance $R$. 
        The P-wave $\rhoz(770)\rhoz(770)$ is a $P$-odd component. The phase change in this
        resonance is tested with the $P$-odd
        \CP-violation test.  Results for all other scenarios are given with
    the standard $P$-even test.}
    \centering
    \begin{tabular}{ll}
        \hline 
        $R$ (partial wave) ($\Delta A$, $\Delta \phi$) & \pvalue (fit) \\
        \hline  
        \decay{a_1}{\rhoz\pi} (S) ($5\%, 0^\circ$) &$2.6^{+3.4}_{-1.7}\times10^{-4}$ \\
        \decay{a_1}{\rhoz\pi} (S) ($0\%, 3^\circ$) & $1.2^{+3.6}_{-1.2}\times10^{-6}$ \\
        \rhoz\rhoz (D) ($5\%, 0^\circ$) & $3.8^{+2.9}_{-1.9}\times10^{-3}$\\
        \rhoz\rhoz (D) ($0\%, 4^\circ$) & $9.6^{+24}_{-7.2}\times10^{-6}$\\
        \rhoz\rhoz (P) ($4\%, 0^\circ$) & $3.0^{+1.2}_{-0.9}\times10^{-3}$\\
        \hline
        \rhoz\rhoz (P) ($0\%, 3^\circ$) & $9.8^{+4.4}_{-3.8}\times10^{-4}$\\
        \hline 
    \end{tabular}
    \label{tab:sensitivity_overview}
\end{table}
\renewcommand{\arraystretch}{1.0}

Various \CP asymmetries are introduced by modifying, for a chosen \Dz flavour,
either the magnitude or the phase of the dominant amplitude contributions: 
\decay{a_1(1260)^+}{\rhoz(770)\pi^+} (in S-wave) and $\rhoz(770)\rhoz(770)$ (both P-wave and D-wave).
The resulting sensitivities are shown in
Table~\ref{tab:sensitivity_overview}. The \pvalues, including their statistical uncertainties,
are obtained from fits of GEV functions.

The asymmetry significances for each simulated event are shown in
Fig.~\ref{fig:method_visualisation}
for $P$-even and $P$-odd \CP-violation tests and projected onto invariant masses of 
the selected three-pion and two-pion subsystems. The \CP violation is introduced as a phase difference 
in either the $a_1(1260)^+$ or P-wave $\rhoz(770) \rhoz(770)$ amplitudes (see Sect.~\ref{sec:Sensitivity}). 
The shapes of the regions with significant asymmetry that are visible
in the one-dimensional projections in
Fig.~\ref{fig:method_visualisation}  cannot be easily interpreted in
terms of the amplitudes and phase differences of the contributing resonances; the two-dimensional spectra are not easy to understand either.
However, repeating this exercise for different scenarios of \CP
violation, the observed features are found to be
sufficiently distinguishable to help identify
the origin of  any \CP asymmetry in the data. 

The sensitivity of the method also depends on the choice of the effect radius $\delta$.
Studies indicate good stability of the measured sensitivity
for values of $\delta$ from $0.3$ to $1\gevgevcccc$, which
are well above the resolution of the $d_{ij}$ and small compared to the size of the phase space.
The value $\delta=0.5\gevgevcccc$ yields the best sensitivity to most of the \CP-violation
scenarios studied and was chosen, prior to the data unblinding, as the default value.
The optimal $\delta$ value may vary with different \CP-violation scenarios.
Hence, the final results are also quoted for values of $0.3\gevgevcccc$ and $0.7\gevgevcccc$.

\section{Systematic effects}
\label{sec:Systematics}

There are two main sources of asymmetry that may degrade or bias the energy test.
One is an asymmetry that may arise from background and the other is due 
to detection and production asymmetries that could vary across phase
space. The effect of these asymmetries is studied for both the $P$-even and $P$-odd \CP-violation measurements.

Background asymmetries are tested by applying the energy test to events in the sidebands  
surrounding the signal region in the $\dm$ vs.~$m(\pip\pim\pip\pim)$ plane. These events are randomly 
split into 11 subsamples containing the same number of background events as expected to contribute under 
the signal peak. The \pvalues obtained are compatible with a uniform
distribution and no significant asymmetry is found; \pvalues range between $2\%$ and $96\%$ ($4\%$  and $91\%$) 
for $P$-even ($P$-odd). 
As the background present in the signal region is found to be \CP
symmetric, no correction is applied in the $T$ value calculation discussed in Sect.~\ref{sec:Method}.
 
Positively and negatively charged pions interact differently with matter; the differences in 
the inelastic cross-sections are momentum dependent and significant for low-momentum particles~\cite{PDG2016}. 
The presence of the \pip\pim pairs in the final state makes the detection asymmetries cancel 
to first order. Residual local asymmetries may remain in certain regions of the phase space where 
\pip and \pim have different kinematic distributions. 
These effects are tested using the Cabibbo-favoured decay \decay{\Dz}{\Km\pip\pim\pip} as a control mode.
This channel is affected by kaon detection asymmetries, which are known to be larger than pion detection 
asymmetries and thus should serve as a conservative test. The data
sample is obtained with the same kinematic selection criteria as for
the signal channel and imposing requirements on the candidate \Kpm particles
identified using information from the ring-imaging Cherenkov
detectors. The control sample is split into ten subsets, each of which contains approximately the same amount of data as the signal sample.
The energy test yields results compatible with a uniform distribution
of \pvalues with values between $3\%$ and $87\%$ ($8\%$  and $74\%$) for $P$-even ($P$-odd), which is consistent with the assumption 
that this source of  asymmetry is below the current level of sensitivity.

Asymmetries in the soft pion detection, although largely uncorrelated with the \Dz phase space, 
are reduced using fiducial cuts (see Sect.~\ref{sec:Selection}). 
Charm mesons produced in the $pp$ interactions exhibit a production
asymmetry up to the percent level, which is 
slightly dependent on the meson kinematics~\cite{LHCb:2012fb,Aaij:2012cy}. 
More \Dstarm particles are observed than \Dstarp, giving rise to a
global asymmetry, to which the applied method is insensitive by construction.
No significant local \CP asymmetry is expected owing to the small observed correlation of the \Dstar momentum and the \Dz phase space.

\section{Results and conclusions}
\label{sec:Results}

\begin{figure}[htb!]
 \centering
 \includegraphics[width=0.39\textwidth]{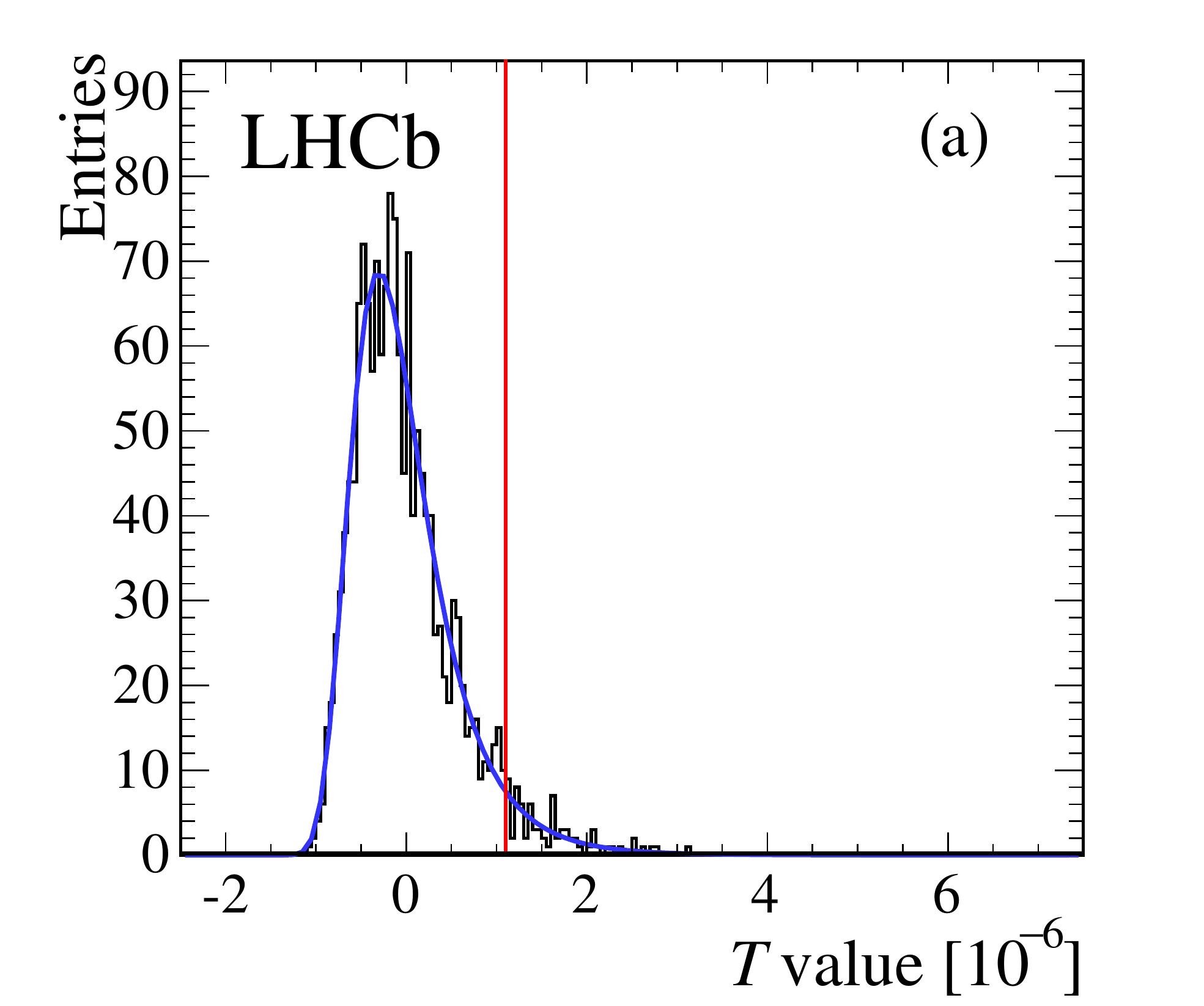}
    \hspace{1cm}
  \includegraphics[width=0.39\textwidth]{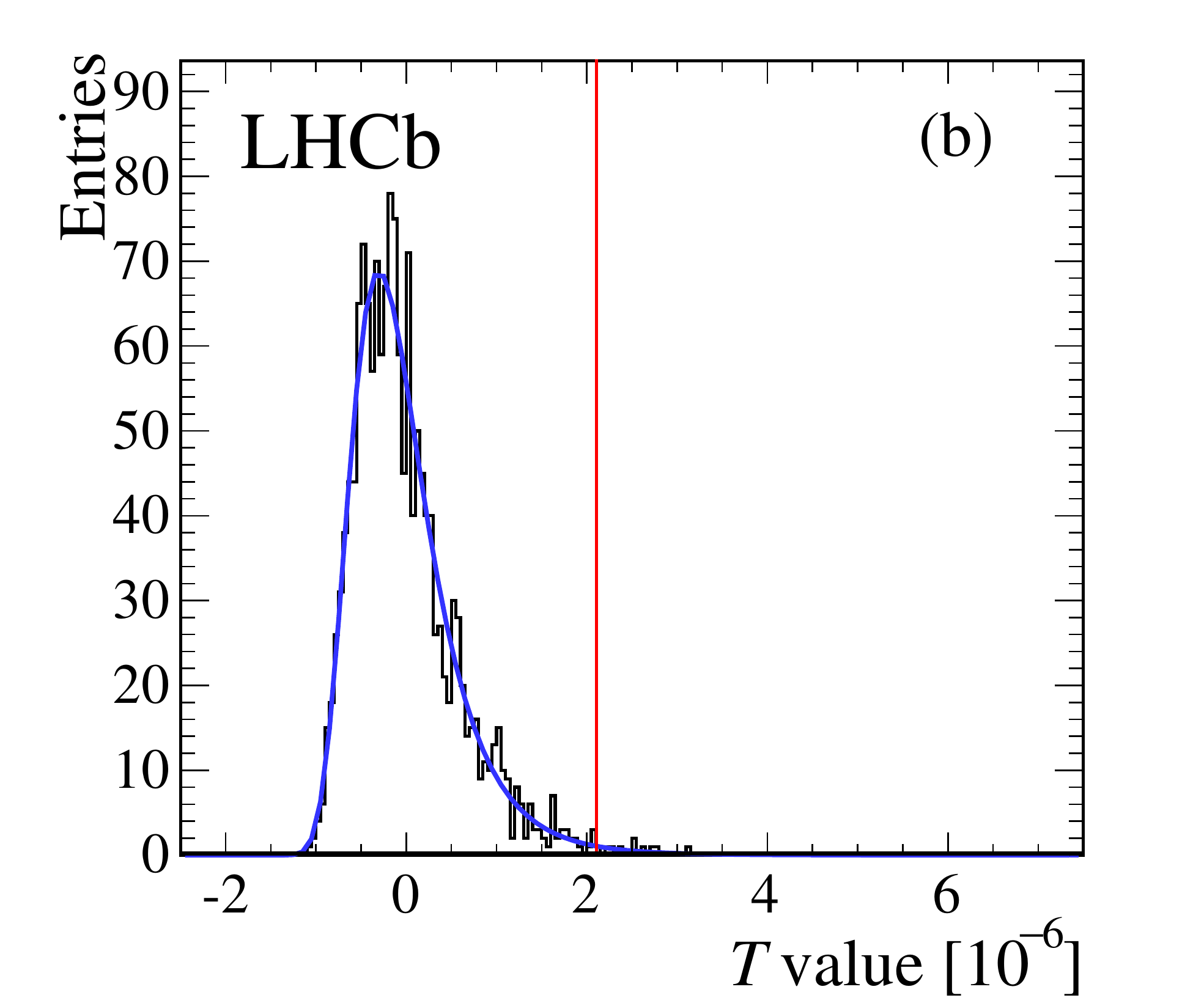}
  
 \includegraphics[width=0.39\textwidth]{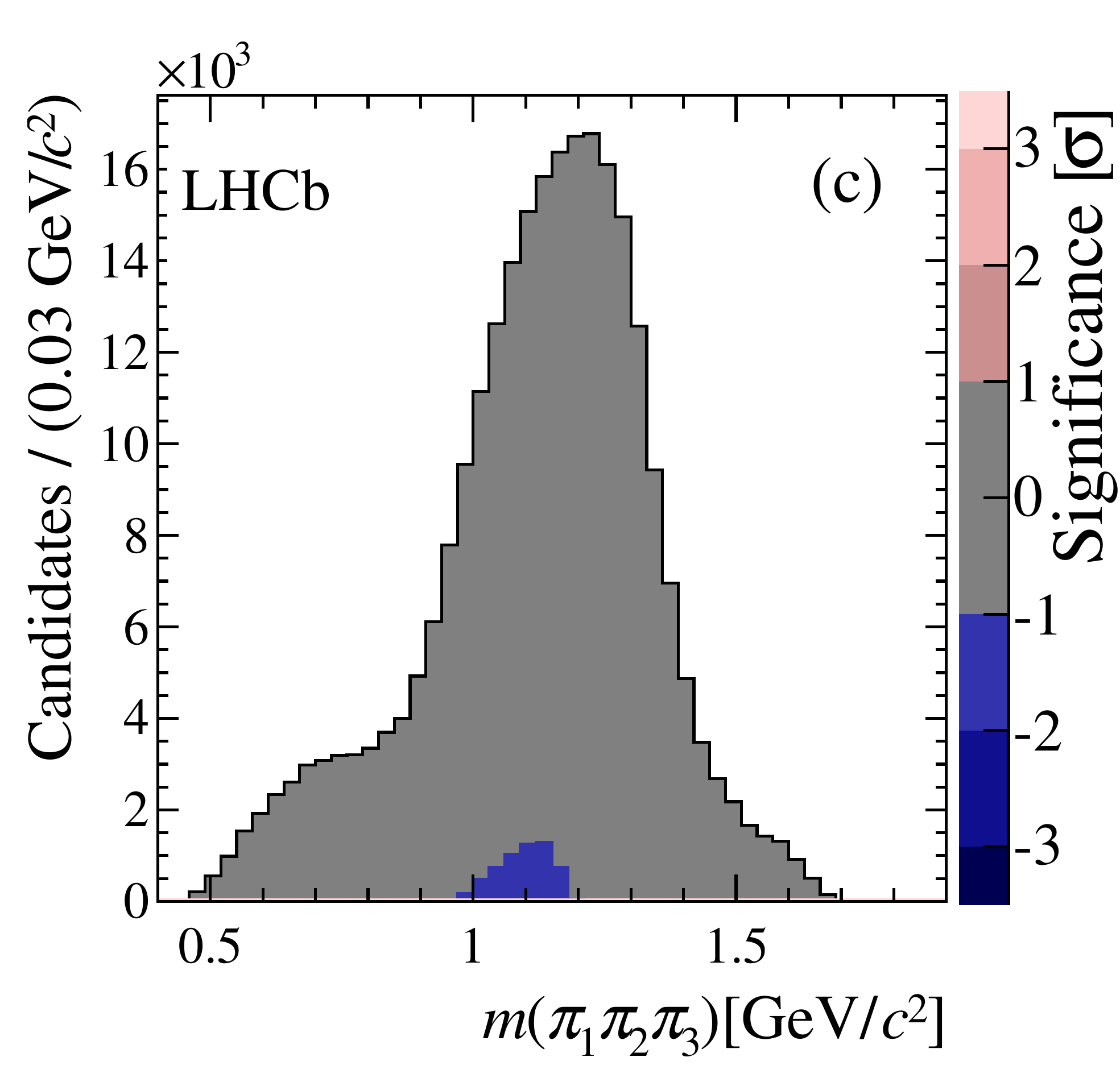}
    \hspace{1cm}
 \includegraphics[width=0.39\textwidth]{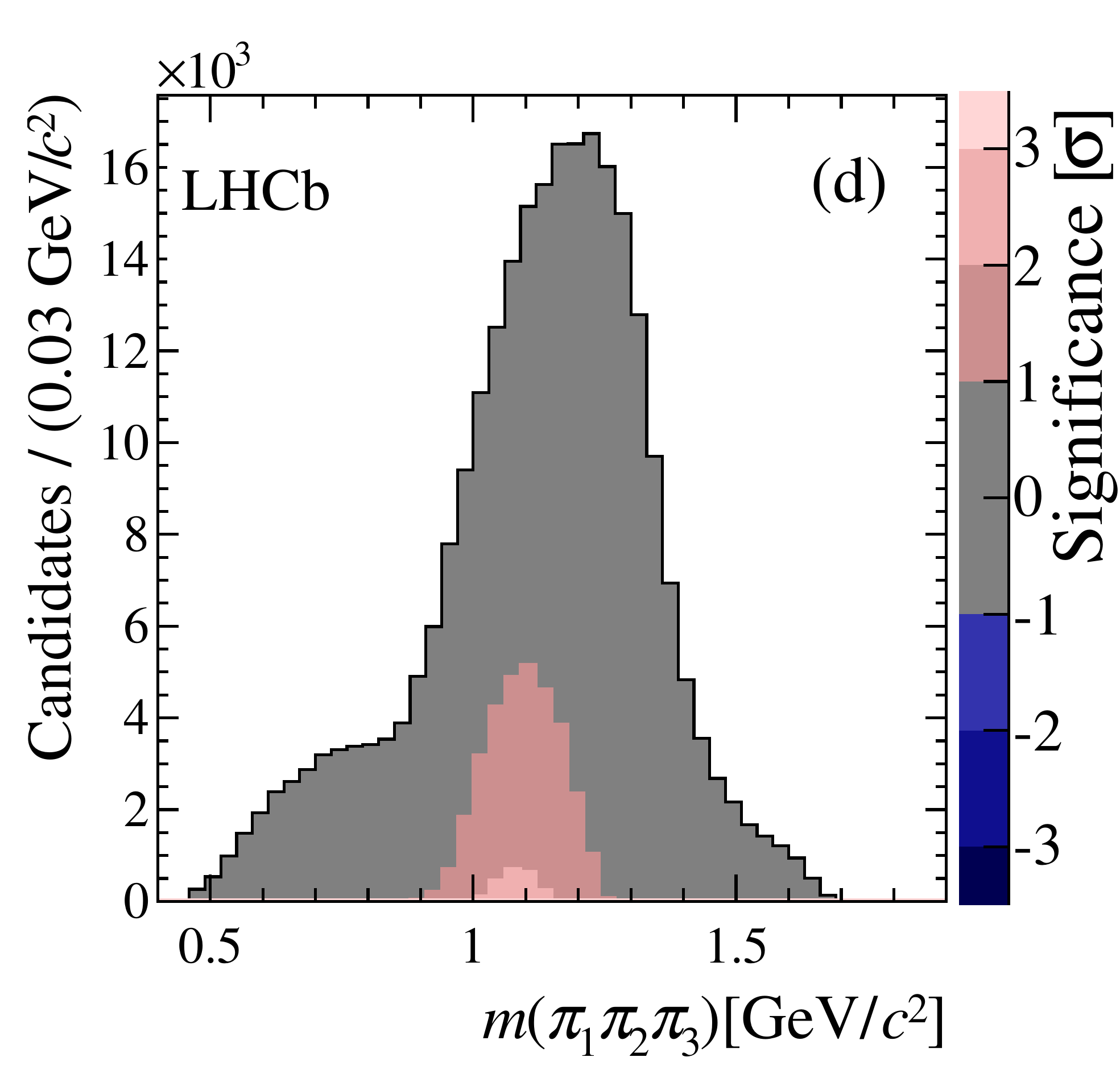}
 
 \includegraphics[width=0.39\textwidth]{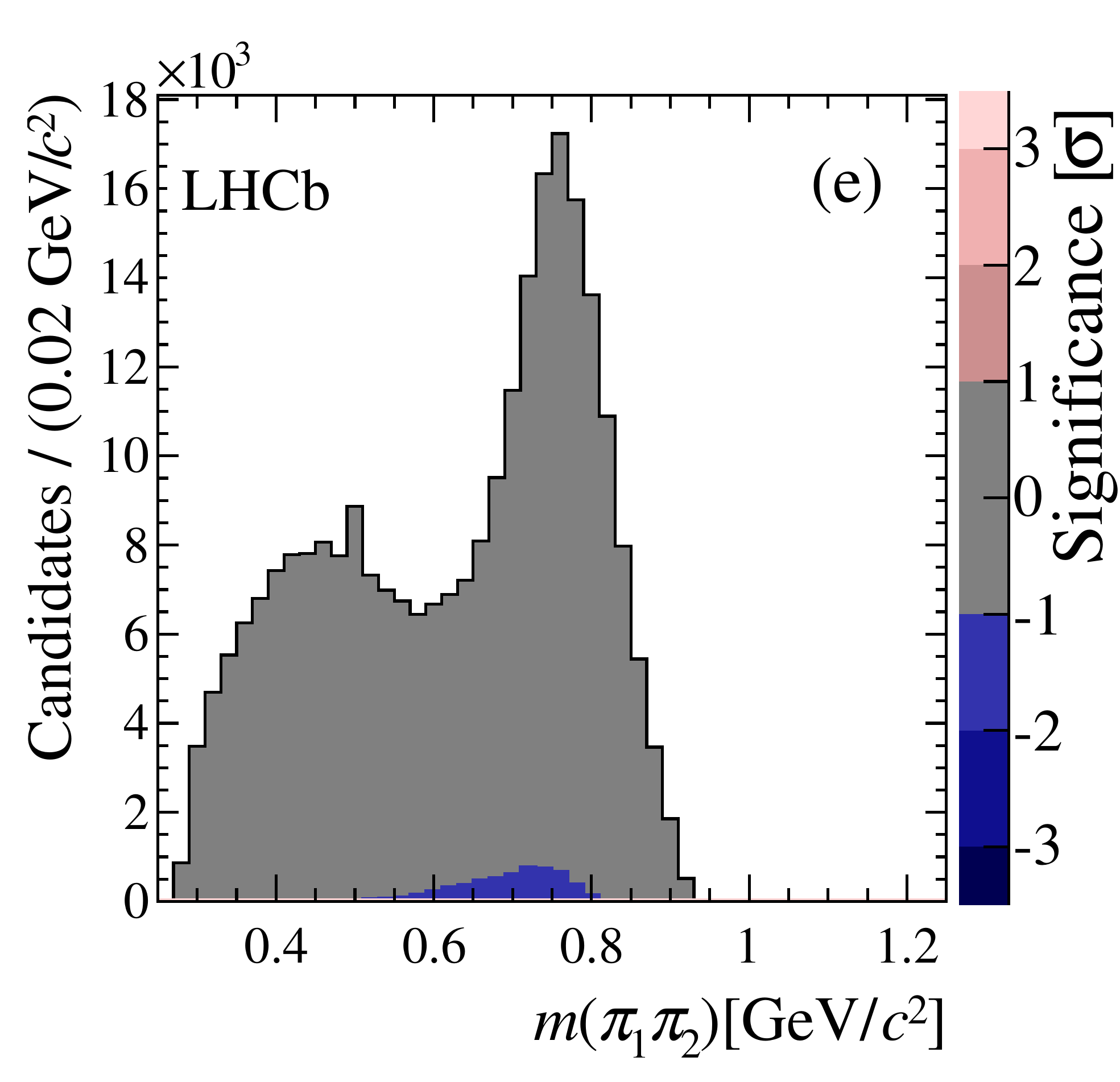}
    \hspace{1cm}
 \includegraphics[width=0.39\textwidth]{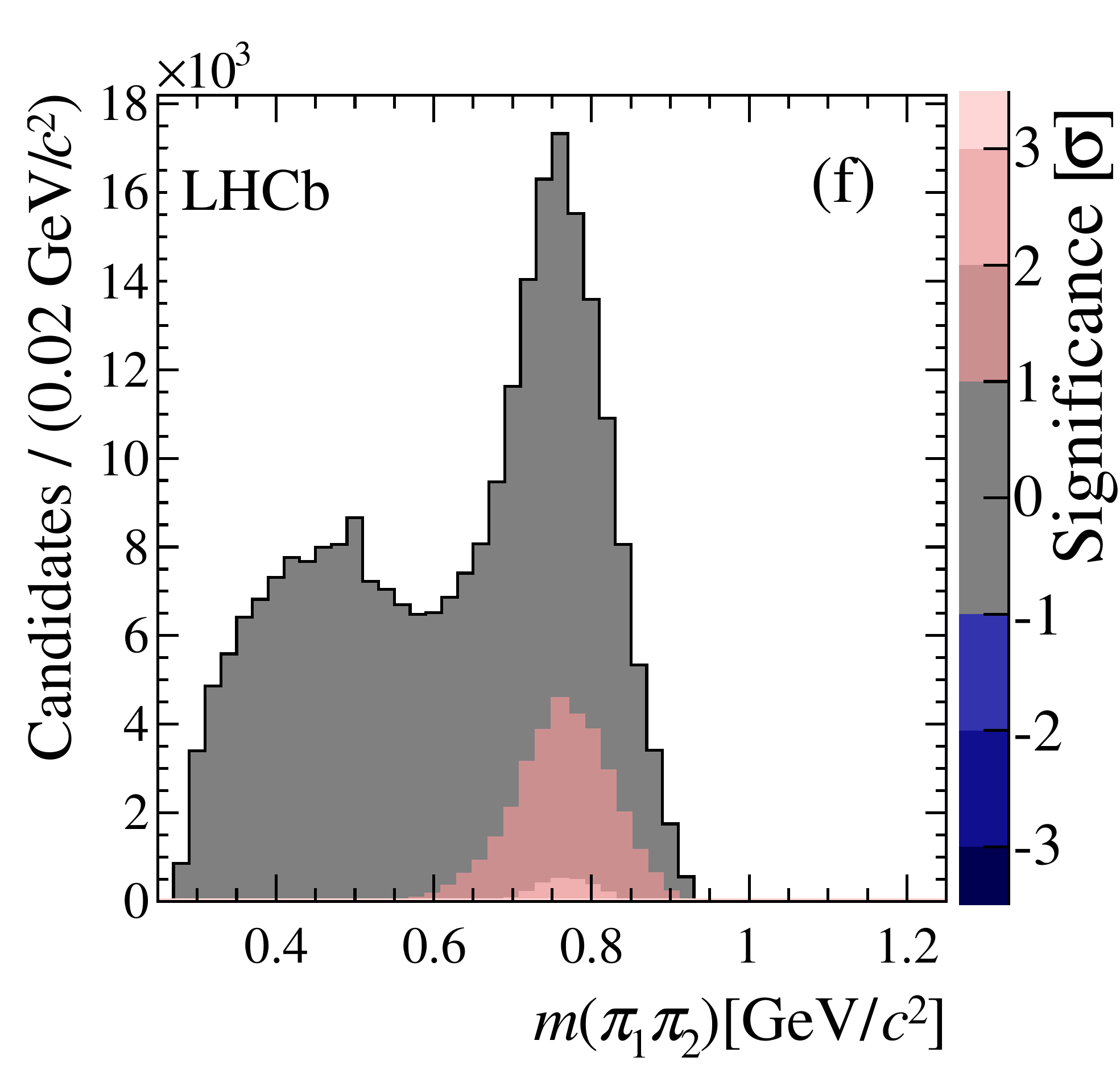}
 \caption{
 \small
       (a,b) Distribution of permutation $T$-values fitted with a GEV function and showing
       the $T$-value of the data tests as a vertical line, and (c,d,e,f) local asymmetry
       significances. Left column plots are for the $P$-even \CP-violation test, projected onto the (c) $m(\pi_1 \pi_2 \pi_3)$
       and (e) $m(\pi_1\pi_2)$ axes. Right column plots are for the
       $P$-odd \CP-violation test projected onto the same axes. In plots
       (c,d,e,f)  the grey area correspond to candidates with a
       contribution to the $T$-value of less than one~standard
       deviation.  In the $P$-even \CP violation test the positive
       (negative) asymmetry significance is set for the \Dz candidates
       having positive (negative) contribution to the  measured $T$
       value.  In the $P$-odd \CP violation test the positive
       (negative) asymmetry significance is set for sample $I+IV$
       having positive (negative) contribution to the measured $T$
       value (see Sect.~\ref{sec:Method}).
      The pink (blue) area corresponds to candidates with a
        positive (negative) contribution to the $T$-value. Light, medium
        or dark shades of pink and blue correspond to 
       between one and two, two and three, and more than
        three standard deviation contributions, respectively.
 \label{fig:results}}
\end{figure}

The application of the energy test to all selected
\decay{\Dz}{\pip\pim\pip\pim} candidates using an effective radius of
$\delta=0.5\gevgevcccc$ yields $T=1.10\times10^{-6}$ in the $P$-even \CP-violation test and 
$T=2.11\times10^{-6}$ in the $P$-odd \CP-violation test.
The permutation $T$ value distributions are shown in Fig.~\ref{fig:results} (a,b).
By counting the fraction of permutations with a $T$ value above the nominal $T$ value in the data,
a \pvalue of $(4.6\pm0.5)\%$ is obtained in the $P$-even \CP-violation test 
and $(0.6\pm0.2)\%$ in the $P$-odd \CP-violation test. The central value of the $P$-odd result would correspond to being at
least 2.7 standard deviations from the mean of a normal distribution.
The significance levels of the $T_i$ values are shown in
Fig.~\ref{fig:results} (c,e) projected onto the $m(\pi_1 \pi_2 \pi_3)$ axis, and Fig.~\ref{fig:results} (d,f) projected onto 
the $m(\pi_1 \pi_2)$ axis.
In the $P$-even test, a small phase space region contains candidates with
a local negative asymmetry exceeding $1\sigma$ significance. 
Furthermore, in the $P$-odd test, 
 candidates with a local positive asymmetry exceeding $2\sigma$ significance are seen in a small phase-space region
dominated by the $\rho^0$ resonance,  which can be compared with the corresponding plots in Fig.~\ref{fig:method_visualisation}.  
Varying the effective radius
results in the \pvalues listed in Table~\ref{tab:results}. 
The central values of the $P$-odd results for $\delta=0.3\gevgevcccc$ correspond to more than three standard deviations from the mean of a normal distribution but the significance falls below this level when considering their uncertainties.

\begin{table}[t!]
\caption{Results for the $P$-even and $P$-odd \CP-violation tests for three different values of the effective radius
  $\delta$. The \pvalues obtained with both the counting and GEV fitting
  methods are given (see text). The counting method is the default method.}
\centering
\begin{tabular}{ccccc}
\hline 
& \multicolumn{2}{c}{\pvalue $P$-even} &  \multicolumn{2}{c}{\pvalue
                                                        $P$-odd } \\
$\delta$~[\gevgevcccc] & Counting & GEV fit & Counting & GEV fit \\
\hline  

0.3 & $(0.88 \pm 0.26)\%$  & $(0.78 \pm  0.10)\%$ 
      & $(0.24 \pm 0.14)\%$  & $(0.28 \pm 0.04)\%$ \\
0.5 & $(4.6 \pm 0.5)\%$      & $(4.8 \pm 0.3)\%$
      & $(0.63 \pm 0.20)\%$  & $(0.34 \pm 0.05)\%$ \\
0.7 & $(16 \pm 2)\%$    & $(17\pm 2)\%$
      & $(0.83 \pm 0.48)\%$  & $(0.52\pm 0.16)\%$ \\

\hline  
\end{tabular}
\label{tab:results}
\end{table}
 
In summary, a search for time-integrated \CP violation in the Cabibbo-suppressed
decay \decay{\Dz}{\pip\pim\pip\pim} is performed using a novel  unbinned
model-independent  technique.  This is the first application of the energy test to four-body decays and extends the approach to allow $P$-odd
\CP violation searches. This analysis has the best sensitivity from a single experiment to
$P$-even \CP violation and is the first test for $P$-odd
\CP violation in this decay.
 The data are found to be marginally
consistent with the hypothesis of \CP symmetry at the current level of precision.

\section*{Acknowledgements}

\noindent We express our gratitude to our colleagues in the CERN
accelerator departments for the excellent performance of the LHC. We
thank the technical and administrative staff at the LHCb
institutes. We acknowledge support from CERN and from the national
agencies: CAPES, CNPq, FAPERJ and FINEP (Brazil); NSFC (China);
CNRS/IN2P3 (France); BMBF, DFG and MPG (Germany); INFN (Italy);
FOM and NWO (The Netherlands); MNiSW and NCN (Poland); MEN/IFA (Romania);
MinES and FASO (Russia); MinECo (Spain); SNSF and SER (Switzerland);
NASU (Ukraine); STFC (United Kingdom); NSF (USA).
We acknowledge the computing resources that are provided by CERN, IN2P3 (France), KIT and DESY (Germany), INFN (Italy), SURF (The Netherlands), PIC (Spain), GridPP (United Kingdom), RRCKI and Yandex LLC (Russia), CSCS (Switzerland), IFIN-HH (Romania), CBPF (Brazil), PL-GRID (Poland) and OSC (USA). This work
was supported in part by an allocation of computing
time from the University of Manchester and the Ohio Supercomputer Center.  
We are indebted to the communities behind the multiple open
source software packages on which we depend.
Individual groups or members have received support from AvH Foundation (Germany),
EPLANET, Marie Sk\l{}odowska-Curie Actions and ERC (European Union),
Conseil G\'{e}n\'{e}ral de Haute-Savoie, Labex ENIGMASS and OCEVU,
R\'{e}gion Auvergne (France), RFBR and Yandex LLC (Russia), GVA, XuntaGal and GENCAT (Spain), Herchel Smith Fund, The Royal Society, Royal Commission for the Exhibition of 1851 and the Leverhulme Trust (United Kingdom).

\appendix
\onecolumn  

\addcontentsline{toc}{section}{References}
\setboolean{inbibliography}{true}
\bibliographystyle{LHCb}
\bibliography{main,LHCb-PAPER,LHCb-CONF,LHCb-DP,LHCb-TDR}

\newpage

\newpage
\centerline{\large\bf LHCb collaboration}
\begin{flushleft}
\small
R.~Aaij$^{40}$,
B.~Adeva$^{39}$,
M.~Adinolfi$^{48}$,
Z.~Ajaltouni$^{5}$,
S.~Akar$^{6}$,
J.~Albrecht$^{10}$,
F.~Alessio$^{40}$,
M.~Alexander$^{53}$,
S.~Ali$^{43}$,
G.~Alkhazov$^{31}$
% ...

{\footnotesize \it
% ...
$ ^{5}$Clermont Universit{\'e}, Universit{\'e} Blaise Pascal, CNRS/IN2P3, LPC, Clermont-Ferrand, France\\
$ ^{6}$CPPM, Aix-Marseille Universit{\'e}, CNRS/IN2P3, Marseille, France\\
% ...
$ ^{10}$Fakult{\"a}t Physik, Technische Universit{\"a}t Dortmund, Dortmund, Germany\\
% ...
$ ^{31}$Petersburg Nuclear Physics Institute (PNPI), Gatchina, Russia\\
% ...
$ ^{39}$Universidad de Santiago de Compostela, Santiago de Compostela, Spain\\
$ ^{40}$European Organization for Nuclear Research (CERN), Geneva, Switzerland\\
% ...
$ ^{43}$Nikhef National Institute for Subatomic Physics, Amsterdam, The Netherlands\\
% ...
$ ^{48}$H.H. Wills Physics Laboratory, University of Bristol, Bristol, United Kingdom\\
% ...
$ ^{53}$School of Physics and Astronomy, University of Glasgow, Glasgow, United Kingdom\\
% ...
}
\end{flushleft}

\end{document}